\documentclass[journal=jctcce]{achemso}

\usepackage{graphicx}
\usepackage{dcolumn}
\usepackage{bm}
\usepackage{xspace}
\usepackage[normalem]{ulem}
\usepackage{units}
\usepackage{dcolumn}
\usepackage{paralist}
\usepackage{natmove}
\usepackage{natbib}
\usepackage{sidecap}
\usepackage{amsmath}
\usepackage{amssymb}
\usepackage{amsfonts}
\usepackage{listings}
\usepackage{subfigure}
\usepackage[toc,page]{appendix}
\usepackage{chemfig}

\usepackage{xcolor}

\newcommand{\ddint}[1]{{\ensuremath{\text{d}^3#1\,}}}
\newcommand{\dint}[1]{{\ensuremath{\text{d}#1\,}}}
\newcommand{\matr}[1]{\ensuremath\underline{\mathbf{#1}}}
\newcommand{\vecr}{{\ensuremath{\mathbf{r}}}}

\newcommand{\xtp}{VOTCA-XTP\xspace}
\newcommand{\gaussian}{GAUSSIAN\xspace}
\newcommand{\orca}{ORCA\xspace}
\newcommand{\nwchem}{NWChem\xspace}

\newcommand{\equ}[1]{eq.~\ref{#1}}
\newcommand{\Equ}[1]{Eq.~\ref{#1}}

\newcommand{\fig}[1]{figure~\ref{#1}}

\newcommand{\sect}[1]{sec.~\ref{#1}}

\newcommand{\ket}[1]{{\ensuremath{\vert #1 \rangle}}}
\newcommand{\bra}[1]{{\ensuremath{\langle #1 \vert}}}

\newcommand{\ks}{\ensuremath{\text{KS}}}
\newcommand{\qp}{\ensuremath{\text{QP}}}

\newcommand{\onlinecite}[1]{\cite{#1}}

\newcommand{\newtext}[1]{}

\newcommand{\note}[1]{{#1}}

\lstdefinestyle{dna}{%
    literate={A}{\textcolor{green}{A}}{1}
        {G}{\textcolor{blue}{G}}{1}
        {C}{\textcolor{red}{C}}{1}
        {T}{\textcolor{black}{T}}{1},
    basicstyle=\ttfamily,
}
\newcommand{\DNA}[1]{%
    \lstinline[style=dna]{#1}%
}

 \title{Electronic Excitations in Complex Molecular Environments:
  Many-Body Green's Functions Theory in VOTCA-XTP}

  \author{Jens Wehner}
  \affiliation{Max Planck Institute for Polymer Research, Ackermannweg
    10, D-55128 Mainz, Germany and Eindhoven University of Technology, Department of
    Mathematics and Computer Science \& Institute for Complex Molecular
    Systems, P.O. Box 513, 5600MB Eindhoven, The Netherlands}

  \author{Lothar Brombacher}
  \affiliation{Max Planck Institute for Polymer Research, Ackermannweg
    10, D-55128 Mainz, Germany}

  \author{Joshua Brown} \affiliation{Dept. of Electrical Computer and
    Energy Engineering, University of Colorado Boulder, 425 UCB,
    Boulder, CO, 80309, US \& Renewable and Sustainable Energy
    Institute, University of Colorado Boulder, 4001 Discovery Dr,
    Boulder, CO 80303, US}

  \author{Christoph Junghans}\affiliation{Computer, Computational, and
    Statistical Sciences Division, Los Alamos National Laboratory, Los
    Alamos, New Mexico 87545, USA}

  \author{Onur \c{C}aylak}
  \affiliation{Eindhoven University of Technology, Department of
    Mathematics and Computer Science \& Institute for Complex Molecular
    Systems, P.O. Box 513, 5600MB Eindhoven, The Netherlands}

      \author{Yuriy Khalak}
  \affiliation{Eindhoven University of Technology, Department of
    Mathematics and Computer Science \& Institute for Complex Molecular
    Systems, P.O. Box 513, 5600MB Eindhoven, The Netherlands}

  \author{Pranav Madhikar}
  \affiliation{Eindhoven University of Technology, Department of
    Mathematics and Computer Science \& Institute for Complex Molecular
    Systems, P.O. Box 513, 5600MB Eindhoven, The Netherlands}

  \author{Gianluca Tirimb\`{o}}
  \affiliation{Eindhoven University of Technology, Department of
    Mathematics and Computer Science \& Institute for Complex Molecular
    Systems, P.O. Box 513, 5600MB Eindhoven, The Netherlands}

  \author{Bj\"orn Baumeier}
  \email{B.Baumeier@tue.nl}
  \affiliation{Eindhoven University of Technology, Department of
    Mathematics and Computer Science \& Institute for Complex Molecular
    Systems, P.O. Box 513, 5600MB Eindhoven, The Netherlands}

\begin{document}

  \begin{abstract}
    Many-body Green's functions theory within the $GW$ approximation
    and the Bethe-Salpeter Equation (BSE) is implemented in the
    open-source \xtp software, aiming at the calculation of
    electronically excited states in complex molecular
    environments. Based on Gaussian-type atomic orbitals and making
    use of resolution of identify techniques, the code is designed
    specifically for non-periodic systems. Application to the small
    molecule reference set successfully validates the methodology and
    its implementation for a variety of excitation types covering an
    energy range from \unit[2-8]{eV} in single molecules. Further,
    embedding each $GW$-BSE calculation into an atomistically resolved
    surrounding, typically obtained from Molecular Dynamics, accounts
    for effects originating from local fields and polarization. Using
    aqueous DNA as a prototypical system, different levels of
    electrostatic coupling between the regions in this $GW$-BSE/MM
    setup are demonstrated. Particular attention is paid to
    charge-transfer (CT) excitations in adenine base pairs. It is
    found that their energy is extremely sensitive to the specific
    environment and to polarization effects. The calculated redshift
    of the CT excitation energy compared to a nucelobase dimer treated
    in vacuum is of the order of \unit[1]{eV}, which matches
    expectations from experimental data. Predicted lowest CT energies
    are below that of a single nucleobase excitation, indicating the
    possibility of an inital (fast) decay of such an UV excited state
    into a bi-nucleobase CT exciton. The results show that \xtp's
    $GW$-BSE/MM is a powerful tool to study a wide range of types of
    electronic excitations in complex molecular environments.
\end{abstract}

\maketitle

\section{Introduction}
The functionality of many complex supramolecular assemblies is often
not only defined by structural features and dynamics, but also by
their electronic excitations. Photosynthetic processes or microbial
respiratory
activity~\cite{Logan_Exoelectrogenicbacteriathat_2009,Bond_ElectronTransportGeobacter_2012}
are two prominent cases from nature in which complex processes emerge
only from the interplay between electronic structure of molecular
building blocks, such as a single chromophore, and nano- and mesoscale
morphology~\cite{Pirbadian_ShewanellaoneidensisMR1_2014,Yates_ThermallyActivatedLong_2015}. Designing
synthetic materials with similar functionality is attractive for many
technological applications, such as in organic
electronics\cite{Goushi_Organiclightemittingdiodes_2012,Nakanotani_Highefficiencyorganiclightemitting_2014},
thermoelectricity~\cite{Lu_reviewcarrierthermoelectrictransport_2016},
or in sensing and
spectroscopy~\cite{Bagatolli_LAURDANFluorescenceProperties_2012}.

Understanding and controlling this intimate interplay is crucial for a
rational design of such materials, and computational studies hold an
enormous potential in this context. Explicitly linking electronic and
classical degrees of freedom is required, which on the relevant scales
can only be achieved by coupled quantum-classical
techniques. Approaches are needed that provide a high level of
predictability for different types of excitations and that can be
applied to realistic system sizes. Density functional
theory~\cite{Kohn_SelfConsistentEquationsIncluding_1965,Kohn_NobelLectureElectronic_1999}
(DFT) is commonly used to determine ground state properties with
reasonable accuracy, even for systems of considerable size. Reliable
predictions in supramolecular systems are however often sensitive to
the choice of an exchange-correlation (xc) functional or the addition
of appropriate corrections for van-der-Waals effects. Interpretation
of Kohn-Sham energies as single-particle excitation energies is
subject to significant self-interaction errors and a lack of
accounting for electronic screening effects as a response to the
excitation~\cite{Cai_Failuredensityfunctionaltheory_2002}. Coupled
electron-hole excitations, e.g., after photon absorption, are usually
determined within the framework of time-dependent DFT
(TD-DFT). Quality of results obtained with TD-DFT varies significantly
with the excitation's character. In particular the accurate
description of excitons involving extended $\pi$-systems or ones with
charge transfer character has been notoriously
difficult~\cite{Cai_Failuredensityfunctionaltheory_2002,Dreuw_FailureTimeDependentDensity_2004}.\note{
  Functionals which separate long range and short range
  interactions~\cite{Kummel_ChargeTransferExcitationsChallenge_2017,Stein_ReliablePredictionCharge_2009}
  can alleviate the problem, need to be adjusted to a specific system
  for good accuracy.} The usage of accurate wave-function based
techniques, like CASPT2 (complete-active-space second-order perturbation theory)
\cite{Andersson_Secondorderperturbationtheory_1990,Andersson_Secondorderperturbation_1992} or 
CC (coupled cluster) linear response theory using CC3 or CCSDT\cite{Hald_analysisimplementationgeneral_2001,Koch_Coupledclusterresponse_1990}, is
often prohibitive due to their computational demands in application to
large molecular systems.

Traditionally, many-body Green's functions theory with the $GW$
approximation and the Bethe-Salpeter Equation
(BSE)~\cite{Onida_Electronicexcitationsdensityfunctional_2002} has been used within the solid
state community.Recently, however it has gained attention for the
treatment of electronically excited states of molecular
systems~\cite{Ma_Excitedstatesbiological_2009,Blase_FirstprinciplesGWcalculations_2011,Baumeier_FrenkelChargeTransferExcitations_2012,Marom_BenchmarkGWmethods_2012}. $GW$-BSE has been
shown to yield accurate descriptions of different types of
excitations, such as localized (Frenkel) and bimolecular charge
transfer (CT) excitons, on an equal
footing~\citep{Baumeier_FrenkelChargeTransferExcitations_2012, Blase_FirstprinciplesGWcalculations_2011,
 Sharifzadeh_LowEnergyChargeTransferExcitons_2013}. Its computational cost is comparable
to that of TD-DFT, which makes application to molecules or clusters of
molecules of technological relevance
tractable~\cite{Faber_Excitedstatesproperties_2014,Baumeier_ElectronicExcitationsPush_2014,Bagheri_Solventeffectsoptical_2016}
. With the target of studying electronically
excited states in complex molecular environments in mind, one of the
current challenges is to link $GW$-BSE to atomistic models of large
scale morphologies, thus treating these system in hybrid quantum-classical
(QM/MM) setups.

In this paper, we introduce our Gaussian orbital implementation of
$GW$-BSE and its integration into a polarizable QM/MM workflow in the
open-source \xtp package. We first benchmark our implementation using
the Thiel
set\cite{Schreiber_Benchmarkselectronicallyexcited_2008,Silva-Junior_Benchmarkselectronicallyexcited_2008}
containing 28 small molecules of different types: unsaturated
aliphatic hydrocarbons, aldehydes, ketones, amides, aromatic
hydrocarbons and heterocycles, as well as four nucleobases. The set
offers reference data from both experiment and high-order wave
function techniques for a representative variety of types of
excitations, including $\pi\rightarrow\pi^*$ (e.g. ethene),
$n\rightarrow\pi^*$ (e.g. cyclopropene) and $\sigma\rightarrow\pi^*$
(e.g. pyrazine) excitations. These different excitations also cover a
wide range of energies from \unit[2]{eV} to \unit[8]{eV}. After that,
we illustrate the capabilities of the $GW$-BSE/MM setup by
investigating a water-solvated DNA double-strand as a prototypical
system.

Photophysical processes triggered by the absorption of
ultraviolet solar radiation can cause damage and subsequent mutations
in DNA~\cite{Crespo-Hernandez_UltrafastExcitedStateDynamics_2004}. Due to the complexity
of such biological systems, an understanding of the processes
involving excited states is extremely challenging. Bimolecular charge
transfer (CT) excitations between base pairs are considered to play an
important role in the excited-state dynamics in DNA and the mechanism
by which these dynamics lead to structural or chemical decay and,
eventually, gene
mutations~\cite{Cadet_Sensitizedformationoxidatively_2009}. One of
the proposed processes is the rapid decay of an initially photoexcited
$\pi \to \pi^*$ state to a longer-lived CT state, which induces either
structural modifications or chemical oxidation/reduction
reactions~\cite{Kwok_FemtosecondTimeWavelengthResolved_2006}.

As a first step to gain further insight into the exact conditions
under which dynamical excited state processes of this kind can occur
in DNA, a detailed understanding of the CT state energies is
vital. While it is suggested from experiment on water-solvated
single-strands of 20 adenine bases (A$_{20}$) that CT states cause a
faint UV absorption at energies below the energy of the UV active
$\pi\to\pi^*$ transition at approximately
\unit[5]{eV}~\cite{Banyasz_BasePairingEnhances_2011} , a high-level second-order
approximate coupled-cluster method yields CT excitations far above
that for an isolated A$_2$ dimer in the gas
phase~\cite{Lange_BothIntraInterstrand_2009}. Due to the long-range electron-hole
interaction, CT energies are typically very sensitive to the
arrangement of the constituent monomers of the base pair. Optimized
gas-phase structures are likely to exhibit different stacking
distances and even motifs compared to a real single-strand, and lack
the effect of the aqueous environment not only on the structure but
also on the electrostatic environment entirely.

Based on the experimental evidence, one should expect a redshift on
the order of \unit[1]{eV} for CT energies of DNA in an aqueous versus
a vacuum environment. Inclusion of a polarizable environment using a
Polarizable Continuum Model (PCM) on top of
TD-DFT~\cite{Santoro_ExcitedStatesDecay_2009} fails to reproduce this
observation, with the redshift being reported to be as small as
\unit[0.1]{eV}\note{, even using functionals specifically containing
  long-range interactions. The use of model structures consisting of
  idealized molecular dimers, and the lack of explicit local fields
  from an molecular environment comprising, e.g., a charged backbone,
  water solvent, and ions, and the quality of the functional used in
  the TD-DFT calculation to describe CT excitons are likely origins of
  this discrepancy.} To reliably distinguish the different effects of the
aqueous environment and to quantify how they affect the character of
CT excitations contributing to the observed redshift, it is important to
consider a realistic morphology of aqueous DNA and treat it with an
accurate set of techniques. The $GW$-BSE formalism has been reported
to yield very accurate predictions of CT excitation energies in
prototypical small-molecule dimers. Subsequently, Yin et
al.~\cite{Yin_ChargeTransferExcitedStates_2014} studied small complexes of adenine
dimers and water (A$_2$-(H$_2$0)$_m$). Geometries of the complex were
obtained from Classical Molecular Dynamics, while $GW$-BSE was used to
evaluate the excitation energies. It was found that CT energies are
strongly affected by the dipole electric fields in the first hydration
shell around the A$_2$, giving rise to an overall energetic shift to
below that of the $\pi \to \pi^*$ transition in single adenine, much
more in line with the experimental observation.

Inspired by these results, we will use $GW$-BSE within the QM/MM
framework discussed above on aqueous DNA. We go beyond the model used
by Yin et al~\cite{Yin_ChargeTransferExcitedStates_2014} and instead of a single
hydrated adenine dimer, we consider a full double-strand of DNA
solvated in explicit water. This will allow us to study, among other
things, the effects of realistic stacking, the differences between
intra- and inter-strand charge transfer excitations, and the explicit
effect of the DNA backbone. Given the sensitivity of the CT
excitations to water hydration, inclusion of these structural
parameters can have a substantial influence since the electrostatic
environment will be vastly different from the idealized situation of a
hydrated dimer. We will also be able to study dimers formed by
different types of nucleobases on equal footing. We will further
consider different embedding variants, including vacuum calculations
as well as $GW$-BSE/MM calculations with static and additional
polarizable interactions.  Analyzing these different scenarios makes
it possible to disentangle effects of geometric structure of the
dimer, local electric fields or the structure of the environment, and
electronic polarization.

This paper is organized as follows: In \sect{sec:theory:gwbse}, the
theoretical concept of $GW$-BSE for the calculation of one- and
two-particle excitations is briefly outlined, as well as the idea of
coupling to a classical atomistic environment. Detail of the
implementation in \xtp are given in \sect{sec:implementation}. The
results for the Thiel set and the aqueous DNA system are discussed
in~\sect{sec:results}. A brief summary concludes the paper.


\section{Electronically excited states via $GW$-BSE: A theoretical Framework}
\label{sec:theory:gwbse}
In the following section, major concepts behind $GW$-BSE are
summarized. In order to keep the notation simple, we restrict the
discussion to a spin singlet, closed shell system of $N$
electrons. Its ground state, $\ket{N,0}$, can be calculated using DFT by
solving the \emph{Kohn-Sham} (KS)
equations~\cite{Onida_Electronicexcitationsdensityfunctional_2002}
\begin{equation}
  [T_{0} + V_\text{ext} + V_\text{Hartree}
  +V_\text{xc}]\ket{\phi^\ks_{i}} = \varepsilon^\ks_{i}
  \ket{\phi^\ks_{i}}
  \label{equ:theory:kse}
\end{equation}
in which $T_0$ is the kinetic energy, $V_\text{ext}$ an external
potential, $V_\text{Hartree}$ the Hartee potential, and $V_\text{xc}$
the exchange-correlation potential.

\subsection{One particle excitations}
\label{sec:gwbsetheory}
Particle-like excitations, quasiparticles (QP), in which an electron
is added $(N\rightarrow N+1)$ to or removed $(N\rightarrow N-1)$ from
 $\ket{N,0}$, are described by the
one-body Green's function~\cite{Sham_ManyParticleDerivationEffectiveMass_1966,Hedin_EffectsElectronElectronElectronPhonon_1970}
\begin{equation}
G_{1}(1,2) = -i \bra{N,0} T(\psi(1)\psi^{\dagger}(2) ) \ket{N,0} .
\label{equ:theory:g1_definition}
\end{equation}
In this notation time and space variables are combined into a single
variable (e.g $\left({\vecr_1,t_1}\equiv 1\right)$), T is the time
ordering operator and $\psi$ and $\psi^{\dagger}$ are
the annhilation and the creation electron field operator, respectively.

$G_1$ obeys a Dyson-type equation of motion, which in spectral
representation is:
\begin{equation} \label{equ:theory:eqm}
  [ H_0 + \Sigma(E) ] G_1(E) = E G_{1} (E)\, ,
\end{equation}
where $H_0 = T_{0} + V_{\text{ext}} + V_{\text{Hartree}}$, is the DFT
Hamiltonian in the Hartree approximation, while the
exchange-correlation effects are described by the electron self-energy
operator $\Sigma(E)$. Knowing its exact form is crucial to solve the
problem. It can be shown that ~\equ{equ:theory:eqm} is part of a
closed set of coupled equations, known as \emph{Hedin
  equations}~\cite{Hedin_NewMethodCalculating_1965,Strinati_ApplicationGreenfunctions_1988}. DFT,
for instance, can be seen as an approximated solution for the excited
electrons problem, in which $\Sigma \sim V_\text{xc}$. The related
Green's function $G_{0}$, solution of \equ{equ:theory:eqm}, is
$G_{0}(E)=\sum_{i}\frac{\ket{\phi^\ks_{i}}\bra{\phi^\ks_{i}}}{E-\varepsilon_{i}^\ks
  \pm \imath \eta}$.

An approximated solution beyond DFT of this system is given by the
\emph{GW approximation}, in which
\begin{equation}
  \Sigma=iG_1W\, ,
  \label{equ:theory:pentaa}
\end{equation}
where $W=\epsilon^{-1} v_{c}$ is the screened Coulomb interaction and
$v_{c}(\vecr,\vecr') = 1 / |\vecr-\vecr'|$ is the bare Coulomb
interaction, respectively. The inverse dielectric function,
$\epsilon^{-1}$, is calculated in the \emph{random-phase
  approximation} (RPA)~\cite{Hybertsen_FirstPrinciplesTheoryQuasiparticles_1985}. Within
this $GW$ approximation,~\equ{equ:theory:eqm} is transformed into a
Dyson equation of motion for the quasiparticles~\cite{Aulbur_QuasiparticleCalculationsSolids_2000,Rohlfing_Excitedstatesmolecules_2000}:
\begin{equation}
  \left[ H_0 +\Sigma(\varepsilon_i^\qp)\right] \ket{\phi^\qp_i}=\varepsilon_i^\qp\ket{\phi^\qp_i} \,,
 \label{equ:gw_eigen}
\end{equation}
where $\varepsilon_i^\qp$ are the one particle excitation energies of
the system (i.e., the QP electron and holes states) and
$\ket{\phi^\qp_i}$ the quasiparticle wave functions.

In practice, these quasiparticle wave functions are expanded in terms
of the KS states according to
$\ket{\phi^\qp_i}=\sum_{j}a_{j}^{i}\ket{\phi^\ks_j}$. Assuming that
$\ket{\phi^\qp_i}\approx\ket{\phi^\ks_i}$, the quasiparticle energies
can be obtained perturbatively as
\begin{equation}
  \varepsilon_i^\qp= \varepsilon_i^\ks + \Delta \epsilon_i^{GW} =
  \varepsilon_i^\ks + \bra{\phi^\ks_i}
  \Sigma(\varepsilon_i^\qp)-V_\text{xc} \ket{\phi^\ks_i} .
  \label{equ:theory:gw_sc}
\end{equation}
Both the correction term $\Delta \varepsilon_i^{GW}$  and the
non-local, energy-dependent microscopic dielectric function calculated
within the RPA depend on $\varepsilon_i^\qp$. Solutions to
\equ{equ:theory:gw_sc} therefore in general need to be found
self-consistently. It can be avoided by setting
$\varepsilon_i^\qp = \varepsilon_i^\ks$ in the evaluation of $\Sigma$
and $W$, typically called the $G_0W_0$ approximation. To improve on
this one shot approach, the $G_0W_0$ results can be used as the first
step of an iterative evaluation of $\Sigma(\varepsilon_i^\qp)$, called
$GW_{0}$. In the $\text{ev}GW$ procedure, the quasiparticle energies
are additionally updated in the RPA calculation, see
also~\equ{equ:gwbse:polar} and~\fig{fig:gwbse:GWBSEworkflow}, until
eigenvalue (ev) self-consistency.

The determination of $\varepsilon_i^\qp$ via ~\equ{equ:theory:gw_sc}
typically holds if the off-diagonal elements of the self-energy, i.e.,
$\bra{\phi^\ks_j}\Sigma(E)\ket{\phi_i^\ks}$, are small. Otherwise,
expressing the QP wave functions as a linear combination of KS states
need to be fully taken into account. Quasiparticle wave functions and
energies can then be obtained by diagonalizing the energy dependent QP
Hamiltonian
\begin{equation}
H_{i,j}^\qp (E) = \varepsilon^\ks_{i} \delta_{i,j} +
\bra{\phi^\ks_{i}} \Sigma(E)-V_\text{xc} \ket{\phi^\ks_{j}} .
\label{equ:theory:qphamiltonian}
\end{equation}
\note{The $GW$ approach in which also the resulting quasiparticle wave functions, and not only
  the energies, are fed back into the RPA~\equ{equ:gwbse:polar} is
  also referred to as self-consistent $\text{sc}GW$.}

\subsection{Two-particle excitations}
\label{sec:bsetheory}
Neutral excitations, in which the number of electrons remains the same
but they assume an excited configuration $S$
($\ket{N,0} \rightarrow \ket{N,S}$), can be described based on the two
particle Green's function~\cite{Hedin_EffectsElectronElectronElectronPhonon_1970}. It can be
obtained solving a Dyson-like equation of motion, known as the
\emph{Bethe-Salpeter-Equation}
(BSE)~\citep{Strinati_ApplicationGreenfunctions_1988}. Defining the electron-hole
correlation function as
\begin{equation}\label{equ:theory:Lcorrfunc}
 L(12,1'2')=-G_2(12,1'2')+G_1(12)G_1(1'2')\, ,
\end{equation}
where the second term represents the independent movement of two
particles (i.e electron and hole ) as a product of single particle
Green's functions and $G_2$ as the two-particle Green's function. The
BSE reads
\begin{equation}
\begin{aligned}
 L(12,1'2')= & L_0(12,1'2')+\int \dint{(3456)}L_0(14,1'3) \\
 &\times K(35,46)L(62,52')
 \label{equ:theory:G2eqm}
\end{aligned}
\end{equation}
with $K(35,46)$ the interaction kernel and
$L_0(12,1'2')=G_1(1,1')G_1(2,2')$ is the two particle non-interacting
correlation function.

Under the assumption of optical excitations, which involve the
simultaneous creation and annihilation of quasiparticles, we can
reduce the four time variables to two. Due to time homogeneity
$L(12,1'2')$ can be reduced to $L(12,1'2';\omega)$, with the indices
only representing position.  The kernel $K$ is given by the functional
derivative of the full self-energy with respect to non-interacting
quasiparticles.
Using the $GW$ approximation and assuming
$\delta W/\delta G_1\approx 0$, e.g., the screening is not influenced
by the excitation, it can be shown that:
\begin{align}
 K(35,46)&=-i\delta(3,4)\delta(5,6)\nu(3,6)+i\delta(3,6)\delta(4,5)W(3,4)\nonumber\\
 &=K^x(35,46)+K^d(35,46).
\end{align}
$K^d$ is called the \textit{direct interaction} and originates from
the screened interaction $W$ between electron and hole and is
responsible for the binding in the electron hole pair. $K^x$
originates from the unscreened interaction $\nu$ and is responsible
for the singlet-triplet splitting. It is denoted \textit{exchange
  interaction}.

$L_0$ can be written, assuming that $G_1$ is fully given by electron
and hole quasiparticles of the system, as a combination of independent
excitations. In position space it reads
\begin{equation}
\begin{split}
L_0(\vecr_1,\vecr_2,\vecr_1',\vecr_2',\omega)=i & \left[   \sum_{v,c} \frac{\phi_c(\vecr_1)\phi_v(\vecr_2)\phi_v^*(\vecr_1')\phi_c^*(\vecr_2')}{\omega-(\varepsilon_c-\varepsilon_v)+i\eta} \right.\\
&\left. - \frac{\phi_v(\vecr_1)\phi_c(\vecr_2)\phi_c^*(\vecr_1')\phi_v^*(\vecr_2')}{\omega+(\varepsilon_c-\varepsilon_v)-i\eta}\right] \,,
\end{split}
\label{equ:theory:indipendentcorrfunc}
\end{equation}
where $v$ runs over all occupied (hole) states and $c$ over all
unoccupied (electron) states.

Defining the \emph{electron-hole amplitude} as
\begin{equation}
\chi_{S}(\vecr,\vecr') = - \bra{N,0} \psi^{\dagger}(\vecr')\psi(\vecr) \ket{N,S}
\end{equation}
allows to rewrite~\equ{equ:theory:Lcorrfunc} as
\begin{equation}\label{equ:theory:Lcorrfunc_ehampli}
\begin{split}
L(\vecr_1,\vecr_2,\vecr_1',\vecr_2',\omega)=i & \left[ \sum_{S}
  \frac{\chi_S(\vecr_1,\vecr_1')\chi_S^*(\vecr_2',\vecr_2)}{\omega-\Omega_S+i\eta}
\right. \\
&\left. - \sum_{S} \frac{\chi_S(\vecr_2,\vecr_2')\chi_S^*(\vecr_1',\vecr_1)}{\omega+\Omega_S-i\eta}\right]
\end{split}
\end{equation}
where $S$ labels the two-particle excitation and $\Omega_S$ is the
corresponding excitation energy.  To practically evaluate the BSE
in~\equ{equ:theory:G2eqm}, all quantities in
~\equ{equ:theory:Lcorrfunc_ehampli} and
~\equ{equ:theory:indipendentcorrfunc} are expressed in terms of a
basis set of single-particle electron and hole states. That is,
introducing
\begin{equation}
  \chi_S(\vecr_1,\vecr_2)=A_{vc}^S\phi_c(\vecr_1)\phi_v^*(\vecr_2)+B_{vc}^S\phi_v(\vecr_1)\phi_c^*(\vecr_2).
  \label{equ:bsewf}
\end{equation}
transforms the BSE into an eigenvalue problem of the form
\begin{equation}
 \matr{H}^\text{BSE}\ket{{\chi_S}}=\Omega_S\ket{{\chi_S}},
\end{equation}
or in the block matrix form:
\begin{equation}
\begin{pmatrix}
                                    H^{\text{res}}&K \\
                                    -K & -H^{\text{res}}
                                   \end{pmatrix}
 \begin{pmatrix}
 A^S\\ B^S\
 \end{pmatrix}
=\Omega_S
\begin{pmatrix}
 A^S\\ B^S\
 \end{pmatrix}.
 \label{equ:theory:bseeigenvalue}
\end{equation}
The matrix elements $H^\text{res}$ and $K$ are given by:
\begin{align}
 H^{\text{res}}_{vc,v'c'}(\omega)&=D_{vc,v'c'}+K^x_{vc,v'c'}+K^d_{vc,v'c'}\\
 K_{cv,v'c'}(\omega)&=K^x_{cv,v'c'}+K^d_{cv,v'c'}\, ,
\end{align}
where
\begin{align}
D_{vc,v'c'}&=(\varepsilon_v-\varepsilon_c)\delta_{vv'}\delta_{cc'}\label{equ:theory:D}\\
K^x_{vc,v'c'}&=\int \ddint{\vecr} \ddint{\vecr'} \phi_c^*(\vecr)\phi_v(\vecr)\nu(\vecr,\vecr')\phi_{c'}(\vecr')\phi_{v'}^*(\vecr')\\
K^d_{vc,v'c'}&=\int \ddint{\vecr} \ddint{\vecr'}
               \phi_c^*(\vecr)\phi_{c'}(\vecr)W(\vecr,\vecr',\omega=0)\nonumber\\
  &\hspace{1cm}\times \phi_v(\vecr')\phi_{v'}^*(\vecr')
               \, .
\label{equ:theory:bseham}
\end{align}
Here it is assumed, that the dynamic properties of $W(\omega)$ are
negligible and the static approximation $\omega=0$ is used, which
reduces the computational cost significantly. This is only valid if
$\Omega_S-(\varepsilon_c-\varepsilon_v)\ll \omega_l$, where $\omega_l$
is the plasmon frequency, which determines the screening properties.

For many systems the off-diagonal blocks $K$ in
~\equ{equ:theory:bseeigenvalue} are small and can be neglected. This
leads to the Tamm-Dancoff
approximation (TDA)~\cite{Fetter_QuantumTheoryManyparticle_2003}

\begin{equation}
 H^{\text{res}} A^S_{\text{TDA}}=\Omega_S^{\text{TDA}} A^S_{\text{TDA}}
 \label{equ:theory:bseeigenvaluetda}
\end{equation}
and the resulting electron-hole amplitude:
\begin{equation}
 \chi_S^\text{TDA}(\vecr_1,\vecr_2)=\sum_{vc} A_{vc,\text{TDA}}^S \phi_c(\vecr_1)\phi_v^*(\vecr_2).
 \label{equ:theory:BSEwftda}
\end{equation}
This approximation halves the size of the the BSE matrix. Additionally,
it helps to reduce triplet
instabilities~\cite{Rangel_assessmentlowlyingexcitation_2017} but especially for small
molecules the error \note{from neglecting the coupling between resonant and anti-resonant part can
be significant~\cite{Ma_Excitedstatesbiological_2009}.}

The spin structure of the BSE solutions depends on the spin-orbit
coupling. If the ground state is a spin singlet state and spin-orbit
coupling is small, the Hamiltonian decouples into singlet and triplet
class solutions, with $H^\text{BSE}_\text{singlet}=D+K^d+2K^x$ and
$H^\text{BSE}_\text{triplet}=D+K^d$.
If spin-orbit coupling is large, the BSE Hamiltonian must be evaluated
using the full spin structure. More complex spin contributions also
arise for open shell systems, where the ground state is not a singlet.


\subsection{GW-BSE/MM}
Excitation energies in complex molecular environments can be obtained
via a QM/MM
procedure~\cite{Risko_quantumchemicalperspectivelow_2011,Lunkenheimer_SolventEffectsElectronically_2013,
  May_CanLatticeModels_2012,Schwabe_PERICC2Polarizable_2012,Varsano_Theoreticaldescriptionprotein_2017}. This
method relies on treating the active subpart of the system
quantum-mechanically (QM) while embedding it into an environment
described at molecular mechanics resolution (MM). \note{Recently,
  discrete static point charge models of a molecular environment have
  been introduced to plane-wave implementations of
  $GW$-BSE~\cite{Varsano_Theoreticaldescriptionprotein_2017,Varsano_Groundstatestructures_2014}. Such
  static classical models do not include environment polarization
  effects. Furthermore, the use of plane waves and the implied
  articifial periodicity for he study of isolated systems, such as
  molecules, is considered less efficient than the use of localized
  basis sets. Gaussian type orbital implementations of $GW$-BSE, on
  the other hand, have recently been coupled to continuum polarization
  models~\cite{Li_Correlatedelectronholemechanism_2017}, which lack
  explicit local electric fields from a complex molecular
  environment. Inclusion of polarization effects has further been
  reported for the $GW$ formalism using polarizable point charge
  models~\cite{Li_CombiningManyBodyGW_2016}.}
In our QM/MM scheme, we employ a distributed atomic multipole
representation, which allows for a general treatment of static
electric field and polarization effects on equal footing. The QM
subpart can be treated, for istance, at GW-BSE level, and molecules in
the MM region are represented by static atomic multipole moments
$Q^a_t$ ~\cite{Stone_Theoryintermolecularforces_1997} where $t$
indicates the multipole rank and $a$ the associated atom in the
molecule. Additionally, each atom can be assigned a polarizability
$\alpha_{tt'}^{aa'}$ which determines the induced moments
$\Delta Q_t^a$ due to the field generated by moment $t'$ of atom
$a'$. The classical total energy of a system in the state $(s)$ (i.e
neutral or excited) composed of $A$ molecules is given by the sum of
the external (electrostatic) and internal (polarization)
contribution~\cite{Stone_Theoryintermolecularforces_1997}
\begin{equation}
\begin{split}
  E_\text{MM}^{(s)} =&
    \frac{1}{2}\sum_{A}\sum_{A'}{(Q_t^{a(s)} + \Delta Q_t^{a(s)} )
                                T_{tu}^{aa'} (Q_u^{a'(s)} + \Delta
                                Q_u^{a'(s)} )}\\
    &+\frac{1}{2}\sum_{A}\sum_{A'}{\delta_{AA'} \Delta Q_t^{a(s)}
      (\alpha^{-1})_{tt'(s)}^{aa'} \Delta Q_{t'}^{a'(s)} } ,
\end{split}
\label{equ:EMM}
\end{equation}
where interactions between the multipole moment $Q_t^a$ and
$Q_{u}^{a'}$ are described by the interaction tensor $T_{tu}^{aa'}$.
\Equ{equ:EMM} follows a variational principle with respect to the
induced moments and a self-consistent procedure iteratively updating
$\Delta Q_t^a$ is required to find the minimum energy. Induced
interactions are modified using Thole's damping
functions~\cite{Thole_Molecularpolarizabilitiescalculated_1981,vanDuijnen_MolecularAtomicPolarizabilities_1998}
to avoid overpolarization.
\note{Unlike Ref.~\cite{Li_CombiningManyBodyGW_2016} in which
  environment polarization effects are included explicitly in the $GW$
  equations as an additional screening contribution, we employ a
  double-level self-consistency cycle.} At iteration step $m$, the
potential generated by the static and induced moments of the MM region
acting on the QM region is added to the external potential in
\equ{equ:theory:kse}, and a self-consistent converged QM calculation
is performed yielding an electron density $\rho^{m,(s)}_\text{QM}$ for
the QM part. \note{Since the excited state density for state $S$ is not
  directly accessible in $GW$-BSE, we calculate it as
  $\rho^S(\vecr)=\rho_\text{DFT}(\vecr)+\rho_e^S(\vecr)-\rho_h^S(\vecr)$,
  with the hole and electron contribution of the exciton to the
  density obtained from the electron-hole wavefunction according to
\begin{align}
  \rho_h^S(\vecr_h)&=\int \text{d}\vecr_e  |\chi_S(\vecr_e,\vecr_h)|^2\\
  \rho_e^S(\vecr_e)&=\int \text{d}\vecr_h  |\chi_S(\vecr_e,\vecr_h)|^2,
\end{align}
where $\chi_S(\vecr_e,\vecr_h)$ is the two paricle amplitude,
introduced in~\equ{equ:bsewf}. At the moment an equivalent of relaxed
excited state densities as in
Ref.~\cite{Ronca_DensityRelaxationTimeDependent_2014} is not
accessible as analytic gradients are not implemented, and their proper
definition considered among the theoretic challenges in
$GW$-BSE~\cite{Blase_BetheSalpeterequation_2018}.}

The energy of the QM region is (DFT for \note{ground}
$s=n$, DFT+$GW$-BSE for excited $s=x$ states)
$E^{m,(s)}_\text{QM}=E_\text{DFT}^{m,(s)} + \delta_{sx} \Omega^m_S$.
Once $\rho^{m,(s)}_\text{QM}$ is obtained, an effective multipole
representation $\left\lbrace\widetilde{Q}^{a}_{t}\right\rbrace$ is
used in the next evaluation of the MM energy in~\equ{equ:EMM}. Since
the QM electron density already contains the polarization response to
the outside field, no atomic polarizabilities are added to the QM
region representation in this step. These effective multipoles are
thus used to determine (self-consistently) new induced dipoles in the
MM region using~\equ{equ:EMM}, treating the whole system
classically.

Obtaining the total energy at step $m$ for the coupled QM/MM system
requires the subtraction of the interaction energy of the QM charge
distribution with the field generated by the total MM multipoles,
already included in $E_\text{QM}$. In this way, double counting is
avoided. The total energy at step $m$ is thus
\begin{equation}
  E_\text{QMMM}^{m,(s)} =  E^{m,(s)}_\text{QM} + E^m_\text{MM}-
  \frac{1}{2} \sum_{B \in \text{QM}} \sum_{B' \in \text{QM}}
  \widetilde{Q}^{b(s)}_{t}T^{bb'}_{tu}\widetilde{Q}^{b'(s)}_{u} .
\end{equation}
The whole procedure is repeated until the change of total energy
$\Delta E_\text{QMMM}^{m,(s)} = \vert E_\text{QMMM}^{m,(s)} -
E_\text{QMMM}^{m-1,(s)}\vert$, as well as those of the individual
contributions is smaller than $\unit[10^{-4}]{eV}$.  As stated before,
this procedure is valid for sytems in the \note{ground} or excited state:
calculating separately $E_\text{QMMM}$ in both cases and subtracting
the two, will give the excitation energy in the polarizable
envirorment. This procedure assumes that the states of interest and,
in particular, their localization characteristics on the QM cluster
are easily identifiable.

The explicit state dependence of the coupled QM/MM system introduces
another difficulty, in particular when excited states via $GW$-BSE are
calculated. The solution of the BSE yields a spectrum of excitations,
which are ordered according to their energy. These states can be
energetically separated or very close, depending on the specific
system. As a consequence, the index of a specific excitation of
interest can vary for different external potentials at the individual
steps of the QM/MM self-consistency procedure. It is therefore
important to be able to identify the electronically excited state of
interest during the calculation. In practice, a filtering of the total
spectrum is employed which selects states according to some predefined
property. Currently, the selectable properties are the oscillator
strength $f$ for optically active excitations, the amount of charge
transferred ($\Delta q$) from one fragment to another for charge
transfer states. For such filtering criteria to be applicable, it is
implicitly assumed that the overall characteristics of the excited
states do not change significantly during the QM/MM calculation.

\begin{figure}[t]
  \centering
  \includegraphics[width=0.75\linewidth]{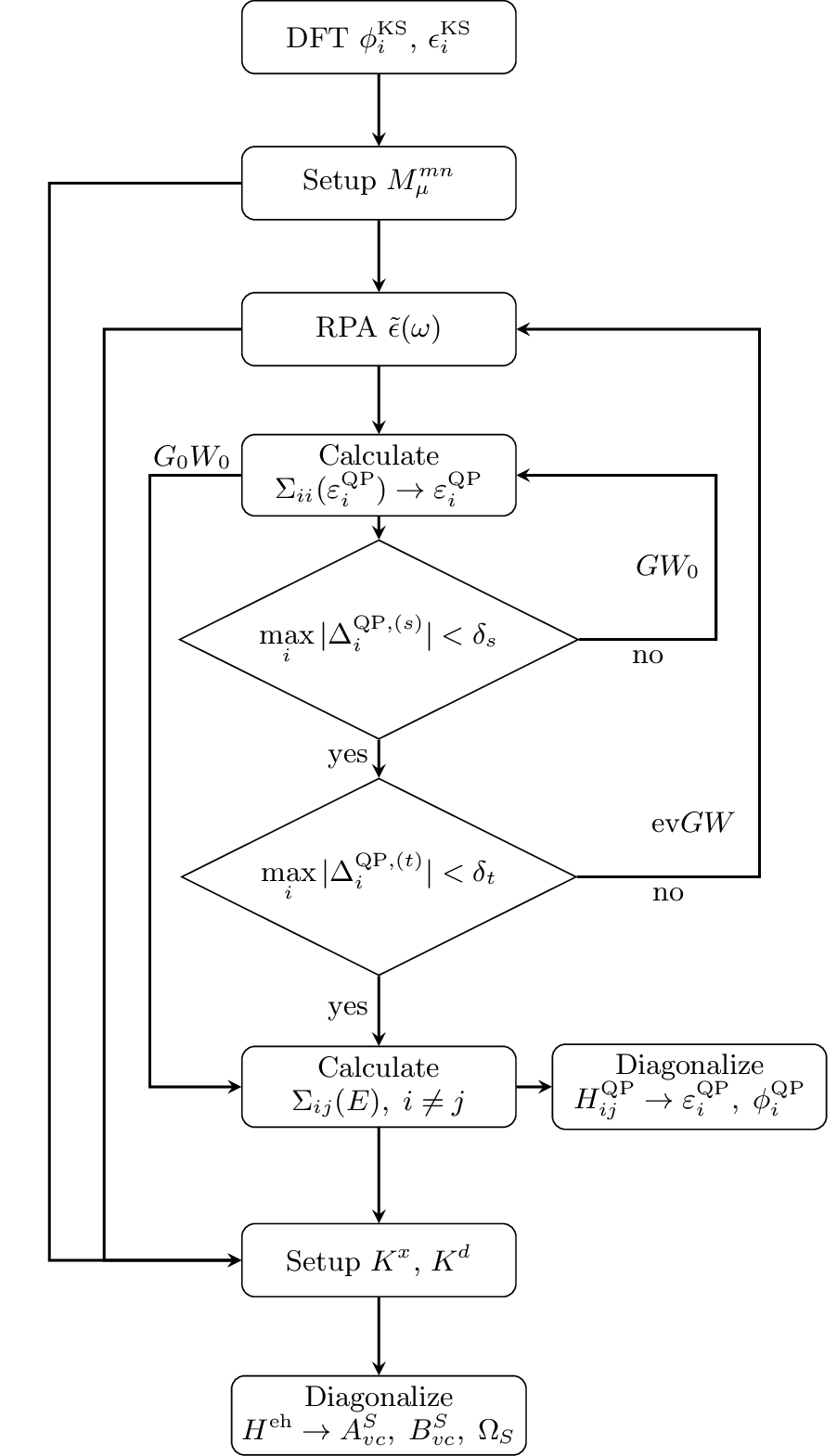}
  \caption[$GW$-BSE workflow as implemented in VOTCA-XTP]{$GW$-BSE
    workflow as implemented in VOTCA-XTP. The inner self-consistency
    loop corresponds to the $GW_0$ algorithm, the outer convergence
    loop, which requires the recalculation of the RPA is the $eVGW$.}
\label{fig:gwbse:GWBSEworkflow}
\end{figure}

\section{Implementation}
\label{sec:implementation}
The theoretical concepts outlined in the previous section are
implemented in the open source \xtp software, available at
\url{github.com/votca}. In the following, we briefly describe the most
important implementational details as they pertain to $GW$-BSE.

Finding the solutions to the quasi-particle equations
\equ{equ:gw_eigen}, and subsequently to the BSE as in
\equ{equ:theory:bseeigenvalue}, requires converged Kohn-Sham molecular
orbitals, their energies, and the contribution of the $V_\text{xc}$ to them
as a starting point. \xtp can read this information from standard
packages using Gaussian-type orbitals (GTOs) as basis functions
$\left\{ \psi_i(\vecr)\right\}$ to express
\begin{equation}
 \phi^\ks_i(\vecr)=\sum_{j=0}^M X_{ij}\psi_j(\vecr)
 \label{equ:theory:basisdecomp}
\end{equation}
and currently provides interfaces to
\gaussian~\cite{Frisch_Gaussian03Revision_2004},
\orca~\cite{Neese_ORCAprogramsystem_2012}, and
\nwchem~\cite{Valiev_NWChemcomprehensivescalable_2010}. The modular nature of the
interfaces allows for straightforward extension to other packages,
provided information about the atomic orbital function order and
input/output files is available. Matrix elements
$\bra{\phi_i^\ks} V_\text{xc} \ket{\phi_j^\ks}$ needed in
\equ{equ:theory:gw_sc} are numerically integrated using spherical Lebedev and
radial Euler-Maclaurin grids as used in
\nwchem~\cite{Valiev_NWChemcomprehensivescalable_2010}, with XC functionals provided by
the \emph{LibXC} library~\cite{Marques_Libxclibraryexchange_2012}.

As an alternative, \xtp also contains a minimal implementation of DFT
with GTOs, which is currently limited to closed shell systems. One-
and two-electron integrals are computed with modified recursive
algorithms~\cite{Obara_Efficientrecursivecomputation_1986,Reine_Multielectronintegrals_2012}. Initial guesses can either be
constructed solving the Hamiltonian of a non-interacting system, or by
superposition of atomic
densities~\cite{VanLenthe_StartingSCFcalculations_2006}. Convergence acceleration
can be achieved by mixing techniques using an approximate energy
functional (ADIIS)~\cite{Hu_Acceleratingselfconsistentfield_2010}, or the commutator of
Fock and density matrix
(DIIS)~\citep{Pulay_ImprovedSCFconvergence_1982}.

The most time-consuming step in a DFT calculation is commonly the
computation of the electron-electron interaction integrals
\begin{equation}
\bra{\psi_i} V_{H} \ket{\psi_j}=\sum_{kl} \matr{D}_{kl} (ij|kl),
\end{equation}
where $\matr{D}$ is the density matrix and
\begin{equation}
 (ij|kl)= \iint \dint{\vecr} \dint{\vecr'}\frac{\psi_i(\vecr)\psi_j(\vecr)\psi_k(\vecr')\psi_l(\vecr')}{|\vecr-\vecr'|}
 \label{equ:gwbse:4c}
\end{equation}
are four-center integrals of Gaussian basis functions. \xtp makes use
of the RI-V approximation, in which the introduction of an auxiliary
basis set with functions $\left\{ \xi_\nu(\vecr)\right\}$ allows to
rewrite \equ{equ:gwbse:4c} as~\cite{Eichkorn_Auxiliarybasissets_1995}
\begin{equation}
 (ij|kl)\approx\sum_{\nu,\mu} (ij|\nu)(\nu|\mu)^{-1} (\mu|kl).
 \label{equ:gwbse:riv}
\end{equation}
Here, $(\nu|\mu)^{-1}$ is the inverse of the two-center repulsion matrix
\begin{equation}
 (\nu|\mu)=\iint \ddint{\vecr_1}\ddint{\vecr_2} \xi_\nu(\vecr_1)\frac{1}{|\vecr_1-\vecr_2|}\xi_\mu(\vecr_2)
 \label{equ:gwbse:2centerrep}
\end{equation}
and $(ij|\nu)$ is the three-center repulsion matrix
\begin{equation}
 (ij|\nu)=\iint \ddint{\vecr_1}\ddint{\vecr_2}\psi_i(\vecr_1)\psi_j(\vecr_1)\frac{1}{|\vecr_1-\vecr_2|}\xi_\nu(\vecr_2).
 \label{equ:gwbse:3centerI}
\end{equation}

The RI-V approximation, sometimes also referred to as density-fitting,
reduces the scaling from $N^4$ to $N^3$, where $N$ is the number of
basis functions, and is particularly useful in application to large
systems. It is also an integral part in the implementation of $GW$-BSE.

Solving the quasiparticle equations~\equ{equ:theory:gw_sc}
or~\equ{equ:theory:qphamiltonian}, relies on the determination of the
self energy $\Sigma$ with the help of the dielectric function within
the RPA. To avoid numerical instabilities in the calculation of the
long range part of $\epsilon$, we introduce a symmetrized (with
respect to $\vecr_1$ and $\vecr_2$) form of the Coulomb interaction
$\tilde{v}_{c}(\vecr_1,\vecr_2)=\pi^{-3/2}/{|\vecr_1-\vecr_2|^2}$,
which is related via the convolution $v_c(\vecr_1,\vecr_2)
  =\int{\tilde{v}_c(\vecr_1,\vecr') \tilde{v}_c(\vecr',\vecr_2)d^3r'}$
  to the actual Coulomb interaction. This can be shown by making use
  of the convolution theorem of Fourier transforms, which yields $v_{c,\mathbf{G}\mathbf{G}'} = \sum_{\mathbf{G}''} \tilde{v}_{c,\mathbf{G}\mathbf{G}''} \tilde{v}_{c,\mathbf{G}''\mathbf{G}'}$. With $\tilde{v}_{c,\mathbf{G}\mathbf{G}'} =
  \delta_{\mathbf{G}\mathbf{G}'}\sqrt{4\pi}/|\mathbf{G}|$ being the Fourier transformed of $\tilde{v}_c(\vecr_1,\vecr_2)$, it follows that $v_{c,\mathbf{G}\mathbf{G}'} =
  \delta_{\mathbf{G}\mathbf{G}'}4\pi/|\mathbf{G}|^2$, which is the known Fourier transform of $v_c(\vecr_1,\vecr_2)$
  (see, e.g., Appendix of
  Ref.~\cite{Rohlfing_EfficientschemeGW_1995} for details). The associated
  symmetrized dielectric function reads $\tilde{\epsilon} = 1
  -\tilde{v}_cP\tilde{v}_c$ or, equivalently, $\tilde{\epsilon} =
  \tilde{v}_c^{-1}\epsilon\tilde{v}_c$. Due to the simultaneous
  symmetrization of the Coulomb interaction and the dielectric
  function, the screened Coulomb interaction is obtained via a double
  convolution as
\begin{equation}
  W = \tilde{v}_c\tilde{\epsilon}^{-1}\tilde{v}_c = \tilde{v}_c
  \left(\tilde{v}_c^{-1}\epsilon \tilde{v}_c  \right)^{-1}\tilde{v}_c
  =\tilde{v}_c  \tilde{v}_c^{-1} \epsilon^{-1} \tilde{v}_c \tilde{v}_c
  = \epsilon^{-1}v_c .
  \label{equ:symmetrization}
\end{equation}
From~\equ{equ:symmetrization} it is apparent that the use of the
symmetrized form of the Coulomb interaction does not change the
screening behavior.

Using the RPA~\citep{Hybertsen_FirstPrinciplesTheoryQuasiparticles_1985,Rohlfing_EfficientschemeGW_1995} for the polarizability
($P=i G_{0}G_{0}$) and the resolution of identity in~
\equ{equ:gwbse:riv}, we can write $\tilde{\epsilon}_{\mu\nu}$ in the
auxiliary basis explicitly as
\begin{equation}
  \begin{split}
 \tilde{\epsilon}_{\mu\nu}(\omega)=\delta_{\mu\nu}+\sum_{m,l}
 I^{ml}_\mu I^{ml}_\nu
 &\left[  \frac{1}{\omega-(\varepsilon_m^\ks-\varepsilon_l^\ks)+i\eta}\right.\\
   &\left. -
     \frac{1}{\omega+(\varepsilon_m^\ks-\varepsilon_l^\ks)-i\eta} \right],
 \end{split}
 \label{equ:gwbse:polar}
\end{equation}
where $I^{ml}_\nu=\sum_\mu (\nu|\mu)^{-1/2} M^{ml}_\mu$ and
$(\nu|\mu)^{-1/2}$ is the matrix square root of the inverse
of~\equ{equ:gwbse:2centerrep}. With the definition of mixed
molecular-atomic three-center Coulomb integrals
\begin{equation}
\begin{split}
  M^{ml}_\mu&=\int \ddint{\vecr_1}\ddint{\vecr_2}
  \phi^\ks_m(\vecr_1)\phi^\ks_l(\vecr_1)\frac{1}{|\vecr_1-\vecr_2|}\xi_\mu(\vecr_2)\\
  &=\sum_{i,j=0}^{i,j=M} X_{im}X_{jl}(ij|\mu)
\end{split}
\label{equ:gwbse:3centerM}
\end{equation}
one obtains the expression of the expectation values of the
self-energy $\Sigma(E)$ with respect to two Kohn-Sham orbitals:
\begin{equation}
  \begin{split}
  \bra{\phi^\ks_n}\Sigma(E)\ket{\phi^\ks_m}=& \frac{i}{2\pi} \sum_{\mu,\nu} \sum_l
  M^{ml}_\mu M^{nl}_\nu \\ &\times \int \dint{\omega}
  e^{i\omega\eta}\frac{\tilde{\epsilon}_{\mu\nu}^{-1}(\omega)}{E-\omega-\varepsilon^\ks_l\pm
    i\eta}.
  \end{split}
 \label{equ:gwbse:selfenergy}
\end{equation}
where the plus (minus) sign in the $\pm i\eta$ term in denominator is
used if the Kohn-Sham orbital $l$ is occupied (empty).

\xtp employs a generalized plasmon pole model (PPM) as outlined in
Ref.~\onlinecite{Rohlfing_EfficientschemeGW_1995} to perform the
frequency integration. This model allows for a quick evaluation of the
integral in~\equ{equ:gwbse:selfenergy}, but at the same time turns the
self-energy into a real
operator~\cite{Rohlfing_ExcitonicEffectsOptical_1998}. \note{The PPM
  was chosen with the application to complex molecular systems of
  considerable size, e.g., with relevance to organic electronics such
  as polymer-fullerene clusters, in mind. The particular model used in
  this work has been successfully applied to determine quasiparticle
  and optical excitations in bulk semiconductor and insulator
  crystals~\cite{Rohlfing_QuasiparticleBandStructure_1995,Rohlfing_Electronholeexcitationsoptical_2000},
  their
  surfaces~\cite{Rohlfing_EfficientschemeGW_1995,Wang_FastInitialDecay_2004},
  defect levels~\cite{Ma_Opticalexcitationdeep_2008}, inorganic
  clusters~\cite{Rohlfing_ExcitonicEffectsOptical_1998},
  polymers~\cite{Rohlfing_OpticalExcitationsConjugated_1999,Artacho_StructuralRelaxationsElectronically_2004,Bagheri_Solventeffectsoptical_2016},
  as well as inorganic and organic
  molecules~\cite{Rohlfing_Excitedstatesmolecules_2000,Ma_Excitedstatesbiological_2009,ma_modeling_2010,Yin_ChargeTransferExcitedStates_2014,Baumeier_FrenkelChargeTransferExcitations_2012,Baumeier_ElectronicExcitationsPush_2014}.}
Explicit integration of the complex integral using (partially)
analytic
techniques~\citep{vanSetten_GWMethodQuantumChemistry_2013,Bruneval_systematicbenchmarkinitio_2015,
  Liu_Numericalintegrationinitio_2015,Rojas_SpaceTimeMethodInitio_1995}
is planned for future versions. The respective matrix elements of the
electron-hole interaction kernel in~\equ{equ:theory:bseeigenvalue} can
be analogously expressed using ~\equ{equ:gwbse:3centerI} and
~\equ{equ:gwbse:3centerM}.

As mentioned in section~\ref{sec:gwbsetheory}, the use of the
Kohn-Sham energies $\epsilon^\ks$ as in ~\equ{equ:gwbse:polar} and
~\equ{equ:gwbse:selfenergy}, corresponds to the so-called single-shot
$G_0W_0$ approximation. In \xtp, the partial self-consistent $GW_0$
and $\text{ev}GW$ schemes are available. \note{Both schemes converge
  the quasiparticle energies $\varepsilon_i^\qp$, but not the
  quasi-particle states $\phi_i^\qp$.  Fully self-consistent $\text{sc}GW$,
  which diagonalises $H_{ij}^\qp$ (\equ{equ:theory:qphamiltonian}) in
  every iteration is currently not implemented.} The respective steps
of the $GW$-BSE calculation are depicted in
\fig{fig:gwbse:GWBSEworkflow}.

The general $GW$-BSE/MM procedure, \note{together with the use of the
  PPM setting \xtp apart from other (closed source) $GW$ codes such as
  Turbomole\cite{Kaplan_QuasiParticleSelfConsistentGW_2016a} or Fiesta\cite{Blase_FirstprinciplesGWcalculations_2011}}, builds on the classical trajectory parsers of
VOTCA-CSG~\cite{Ruhle_VersatileObjectOrientedToolkit_2009} and the
system partitioning functionality and electrostatic and polarizable
interactions and potentials in the VOTCA-CTP
library\cite{Ruhle_MicroscopicSimulationsCharge_2011}.  In addition to
the core functionality described in this paper, \xtp also contains
visualization tools as well as modules for
Mulliken~\cite{Mulliken_ElectronicPopulationAnalysis_1955} or
L\"owdin~\cite{Lowdin_NonOrthogonalityProblemConnected_1950}
population analysis,
CHELPG~\cite{Breneman_Determiningatomcenteredmonopoles_1990} partial
charge fitting for ground, excited, and transition densities with
optional constraints, and numerical excited state gradients and
geometry optimization. It provides methods for the calculation of
non-adiabatic coupling elements for electrons,
holes~\cite{Baumeier_Densityfunctionalbaseddetermination_2010}, and
singlet or triplet
excitons~\cite{Wehner_IntermolecularSingletTriplet_2017}, and links to
a rate-based model of excited state dynamics using kinetic Monte-Carlo
techniques.


\section{Results}
\label{sec:results}
In this section, VOTCA-XTP's $GW$-BSE implementation is first tested
using a small molecule reference set, also known as the {\em Thiel
  set}. After that, we consider as a prototypical complex molecular
system double-stranded DNA with specific focus on the effects of
local-electric fields and environment polarization on charge transfer
excitations.

\begin{figure*}
\centering
  \includegraphics[width=\linewidth]{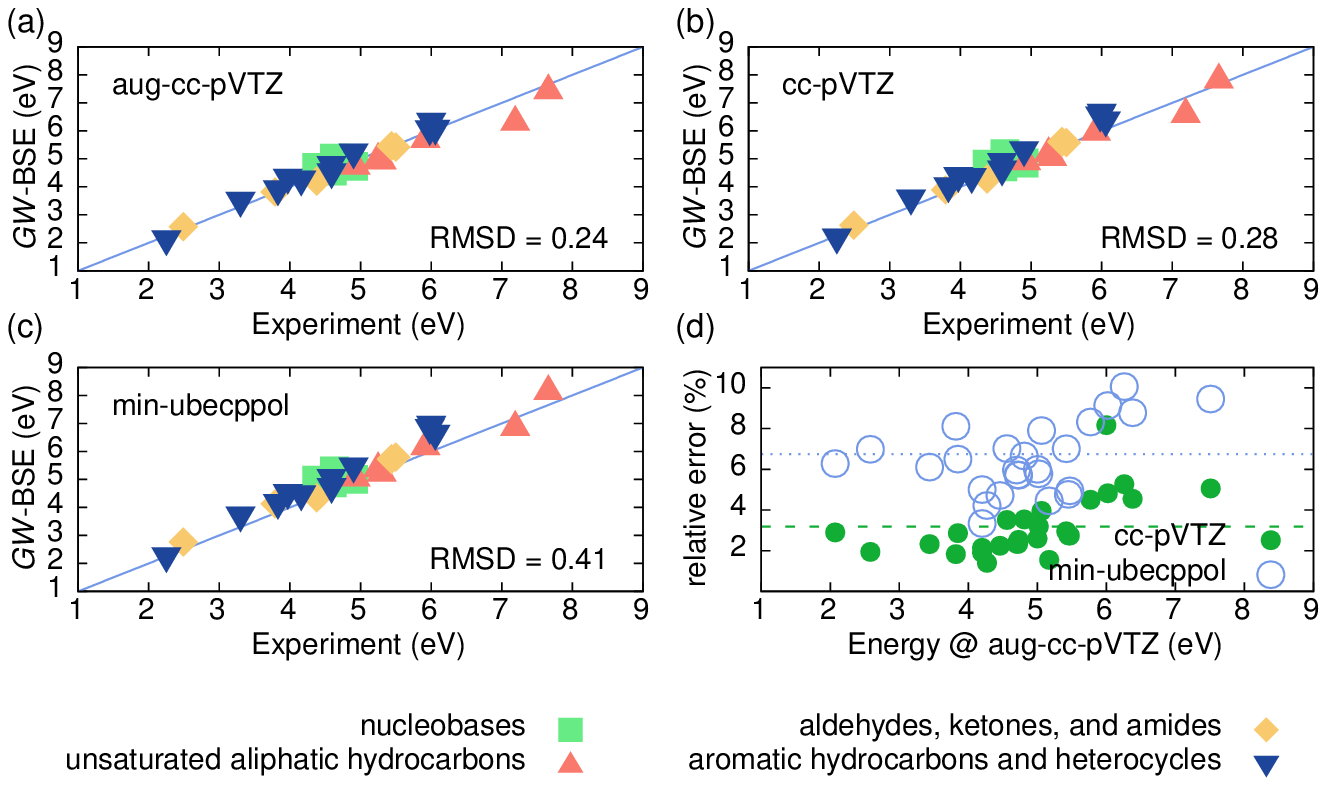}
  \caption{Comparison of calculated lowest singlet excitation energies
    with experimental data (in eV) for the 28 small molecules in
    Thiel's set. Ground state DFT calculations including geometry
    optimizations have been performed on all-electron (AE) level with
    the (a) aug-cc-pVTZ and (b) cc-pVTZ basis sets, as well as (c)
    employing effective core potentials and the min-ubecppol basis
    set, respectively.\note{ The PBE0 functional was used.} The same
    fitting auxiliary basis functions have been used for both DFT and
    $GW$-BSE stages. Data for nucleobases is given by green squares,
    for unsaturated aliphatic hydrocarbons by red up-triangles,
    formaldehydes, ketones, and amides by ocher diamonds, and for
    aromatic hydrocarbons by blue down-triangles. Panel (d) shows the
    relative error between the smaller AE/cc-pVTZ (green filled
    circles) and ECP/min-ubecppol (open blue circles) calculations as
    compared to the more complete AE/aug-cc-pVTZ as a function of
    energy. The green dashed (blue dotted) lines indicate the mean
    error of $\unit[(3.2\pm1.0)]{\%}$ and $\unit[(6.3\pm3.1)]{\%}$,
    respectively. }
  \label{fig:Thiel}
\end{figure*}

\subsection{Single Molecule Data: Thiel set}
For benchmarking, the following procedure has been used. First, the
ground state geometries of the molecules have been optimized using DFT
with the hybrid PBE0
functional~\cite{Adamo_reliabledensityfunctional_1999} at three
different levels of theory, including all-electron (AE) calculations
with the aug-cc-pVTZ and cc-pVTZ basis
sets~\cite{Weigend_Balancedbasissets_2005}, respectively, as well as
calculations making use of effective core potentials and an associated
basis set~\cite{Bergner_initioenergyadjustedpseudopotentials_1993}
that has been augmented by a single shell of polarization functions
taken from the 6-311G**
basis~\cite{Krishnan_Selfconsistentmolecularorbital_1980}. Due to the
significantly reduced computational requirements, the latter case can
be considered a {\em minimal setup} and is further referred to as
min-ubecppol.
%
%
\note{Optimized auxiliary basis sets for
  (aug-)cc-pVTZ~\cite{Weigend_fullydirectRIHF_2002,Weigend_Efficientusecorrelation_2002}
  taken from the Basis Set
  Exchange~\cite{Schuchardt_BasisSetExchange_2007} have been used in
  the resolution-of-identity steps. For the min-ubecppol basis, we
  constructed an auxiliary basis using the technique employed in the
  SAPT
  code~\cite{Misquitta_Intermolecularpotentialsbased_2005,Misquitta_Phthesis_2004,_Generationauxiliarybasis_}. For
  all cases, we have compared the obtained results to those from
  calculations using large auxiliary bases created with the AutoAux
  functionality~\cite{Stoychev_AutomaticGenerationAuxiliary_2017}
  available in Orca~\cite{Neese_ORCAprogramsystem_2012}, and found
  agreement within a few \unit[10]{meV}. A full list of size of the
  corresponding basis sets and auxiliary basis sets is given in
  Tab.~S1 of the Supporting Information.}

For the optimized geometries, excited state energies are determined
within the $GW$-BSE formalism making use of the full BSE
(\equ{equ:theory:bseeigenvalue}) on top of $\text{ev}GW$
self-consistent quasi-particle energies using the procedure outlined
in Sec.~\ref{sec:theory:gwbse}, in which all $GW$ energies are
converged to $\unit[10^{-5}]{Hartree}$. \note{Transitions between all
occupied and empty states, with their total number determined by the
respective basis set sizes as in Tab.~S1 of the Supporting
Information, are taken into account in the calculation of the
dielectric screening in the RPA. This choice is conservatively large,
since including about 10 times as many empty as occupied states has
typically shown to be sufficient to yield converged low energy
excitations (see also Fig.~S1 in the Suporting
Information). Similarly, quasi-particle corrections are determined for
all available states, which are then used to construct the basis of
product states for the expansion of the electron-hole wave functions
in the BSE. For example in the smallest system, ethene, our
calculations with the aug-cc-pVTZ basis include 8 occupied and 130
empty states, leading to 1040 transitions in the RPA and for the BSE
product basis. For naphthalene, inclusion of 34 occupied and 610 empty
states amount to 20740 RPA transitions/BSE product functions.}

From the resulting set of excitations for the respective molecules,
the excitations with optical activity are identified and their
energies are compared to the ones obtained from experiment, as
summarized in Fig.~\ref{fig:Thiel}. The four different categories of
small molecules are represented by differently colored symbols (see
caption for details). For the aug-cc-pVTZ basis set that contains
additional diffuse functions the results depicted in
Fig.~\ref{fig:Thiel}(a) indicate a very good agreement with the
reference data with a RMSD of \note{\unit[0.24]{eV}}. The largest
deviation is found for cyclopropene, whose excitation is reported to
be at \unit[7.19]{eV} in experiment compared to \unit[6.38]{eV} in our
$GW$-BSE calculation. Such a deviation is, however, not unique to our
implementation. In
Ref.~\cite{Jacquemin_BenchmarkingBetheSalpeter_2015}, a $GW$-BSE
excitation energy of \unit[6.14]{eV} was reported, which is very close
to the value of \unit[6.18]{eV} obtained by TD-DFT with the PBE0
functional. Even the {\em Theoretical Best Estimate} based on
high-order wave function methods of \unit[6.65]{eV} shows a similar
deviation. \note{We note that the difference of some of our $GW$-BSE
  results from those in
  Ref.~\cite{Jacquemin_BenchmarkingBetheSalpeter_2015} is likely an
  effect of the different treatment of the frequency dependence of the
  dielectric functions (PPM vs. complex contour integration). Overall
  we find a mean absolute error of \unit[0.14]{eV} between our PPM
  approach and the literature results. For the moderately-sized
  nucleobases, for which one would expect the PPM to be a better
  approximation, this error is as small as \unit[0.03]{eV}. Such an
  error is negligible compared to the effects of the molecular
  environment on the excitation energies, which can be on the order of
  \unit[1]{eV} (see Section~\ref{sect:DNA}).  Smaller
  deviations could also be attributed to more subtle variations in the
  computational protocol, such as in the stabilization of near linear
  dependencies in the basis sets and auxiliary basis sets. In general,
  a more direct comparison between the various theoretical approaches
  is made somewhat difficult by the fact that molecular geometries
  have been optimized at different levels of theory and therefore can
  distort the picture slightly.}

\note{ To scrutinize whether our results are affected by the choice of
  functional in the underlying ground state DFT calculation, we have
  computed the respective excitation energies also with the
  gradient-corrected
  PBE~\cite{Perdew_GeneralizedGradientApproximation_1996} functional
  instead of its hybrid variant PBE0. The full data for both $G_0W_0$
  and $\text{ev}GW$ variants are given in Tab.~S2 in the Supporting
  Information. Inclusion of quasi-particle energie self-consistency
  reduces the mean-absolute error between the PBE0 and PBE functionals
  from $\unit[0.087\pm0.053]{eV}$ to $\unit[0.052\pm0.028]{eV}$. The
  largest difference on $G_0W_0$ level is \unit[0.18]{eV} for
  formaldehyde, compared to only \unit[0.02]{eV} with
  $\text{ev}GW$. Overall, we note only a very weak starting point
  dependence, in particular for $\text{ev}GW$.}

Using diffuse basis functions in quantum-chemical calculations is
typically associated with significant computational costs due to
increased number of functions not only in the basis set itself but
also the auxiliary basis sets for RI. Concomitantly, one occasionally
encounters problems with linear dependencies in the basis sets that
require careful treatment. In this situation, it is desirable to avoid
such diffuse functions, especially in applications to larger
molecules. In Fig.~\ref{fig:Thiel}(b), the $GW$-BSE results obtained
with the cc-pVTZ basis set show overall an excellent agreement with
the experimental reference. \note{On average, the RMSD
of \unit[0.28]{eV} is as expected larger than that for the aug-cc-pVTZ
basis.} This is illustrated in Fig.~\ref{fig:Thiel}(d), in which the
relative deviation of the excitation energies (in \%, indicated by
green filled circles) obtained with cc-pVTZ from those obtained with
the more complete aug-cc-pVTZ basis sets are shown depending on the
absolute aug-cc-pVTZ energies.  It can clearly be seen that on the
energy range covered by the test set, the relative deviation varies
between \unit[1]{\%} and
\unit[9]{\%}, yielding a mean relative error of \unit[3.2]{\%} with
standard deviation of \unit[1.0]{\%}. More importantly, however, the
average run time is reduced to $\unit[(25.2\pm6.5)]{\%}$.

While neglecting diffuse functions already massively reduces
computational costs with only minimal loss of overall accuracy and
reliability, all-electron calculations explicitly include the
typically inert core electrons, such as the two electrons in the $1s$
shell of carbon. It is therefore possible to simply exclude them from
the active space of product functions. However, the presence of such
explicit core electrons requires the use of normal and auxiliary
basis sets with strongly localized functions in the DFT ground
state calculation underlying the $GW$-BSE formalism.

To avoid the expensive calculation of these core states altogether,
effective core potentials can be used in combination with the
min-ubecppol basis set. In Fig.~\ref{fig:Thiel}(c), the obtained
excitation energies are shown compared to the experimental
reference. The overall RMSD of
\note{\unit[0.42]{eV}}, while slightly larger than that recorded for
aug-cc-pVTZ and cc-pVTZ, respectively, is still very good. One can
observe a general tendency for the ECP/min-ubecppol combination to
overestimate the measured data. This is also apparent considering the
relative deviations from aug-cc-pVTZ shown as open circles in
Fig.~\ref{fig:Thiel}(d). Interestingly, the relative deviation varies
between \unit[1]{\%} and \unit[10]{\%}, only slightly larger than for
cc-pVTZ. However the mean error is larger and amounts to
$\unit[(6.7\pm2.0)]{\%}$, which can be considered acceptable, in
particular when one takes into account that the computational cost is
reduced to as much as $\unit[(6.3\pm3.1)]{\%}$ as compared to
aug-cc-pVTZ. These numbers highlight that the use of the minimal
ECP/min-ubecppol variant offers a great compromise between accuracy
and computational cost, which make it particularly attractive for the
application to large, relevant molecular systems.

For completeness, a comparison of the electronic excitation energies
obtained with $GW$-BSE to the Theoretical Best Estimate (TBE) clearly
reveals that all three basis set variants considered in this work
exhibit a very satisfying agreement with the high-order reference. The
data and a figure are available in Fig.~S2 and Tab.~S3 in the
Supporting Information.

Additional savings can in principle be achieved by resorting to the
Tamm-Dancoff Approximation (TDA), in which as explained in
sec~\ref{sec:bsetheory} the resonant-antiresonant coupling terms are
neglected in the Bethe Salpeter Equation. The dimension of the
matrix system is reduced by a factor of two which directly translates
into significant numerical gains. This omission of the corresponding
coupling terms in the BSE can reduce the associated energies by
several \unit[0.1]{eV}, depending on the size of the $\pi$-conjugated
system. The smaller the $\pi$-system, the stronger the effect. For the
relatively small molecules in the Thiel test set, it is therefore
expected that the TDA deviations will be noticeable.

Also for the Thiel set, the TDA energies are typically larger than
those from the full BSE, see Fig.~S3 and Tab.~S3 in the Supporting
Information. Also the size dependence is clearly visible. The
strongest effects can be seen for ethene (C$_2$H$_4$), the molecule
with the smallest $\pi$ system. \note{For the aug-cc-pVTZ basis, TDA
yields an excitation energy of \unit[8.04]{eV} as compared
to \unit[7.51]{eV} obtained by the full BSE
formalism. Resonant-antiresonant coupling accounts for as much
as \unit[0.53]{eV}} in this case. In contrast, for a larger molecule
such as adenine, the effect is reduced to just
\unit[0.02]{eV}. These results illustrate that the TDA can be a useful
approximation depending on the specific system of interest and should
therefore be carefully evaluated.

With these promising conclusions regarding the application to single
small molecule systems at hand, the following section will focus on
the integration of $GW$-BSE in coupled quantum-classical QM/MM setups
for complex molecular environments.


\subsection{Charge transfer excitations in double-stranded aqueous
  DNA}
\label{sect:DNA}
\begin{figure}
\begin{minipage}{\linewidth}
 \centering
  \DNA{GGCGGCGGCGCGGCGTTTTTTGG}\\\DNA{CCGCCGCCGCGCCGCAAAAAACC}
\end{minipage}
\caption{DNA strand sequence.}
\label{fig:DNA_seqeunce}
\end{figure}
To obtain the atomistic structural information, an exemplary DNA
double strand with 23 base pairs in the sequence shown in
Fig.~\ref{fig:DNA_seqeunce} was prepared. This double strand was
solvated by 42216 water molecules and 44 sodium counter ions. For this
system, in the following referred to as aqDNA, classical molecular
dynamics simulations were performed using the AMBER99
forcefield~\cite{Wang_Howwelldoes_2000} for DNA and sodium, and the SPC/E
water model~\cite{Berendsen_missingtermeffective_1987}. Geometric mixing rules
[$\sigma_{ij}=(\sigma_{ii}\sigma_{jj})^{\frac{1}{2}}$ and
$\epsilon_{ij}=(\epsilon_{ii}\epsilon_{jj})^{\frac{1}{2}}$] for
Lennard-Jones (LJ) diameters ($\sigma$) and LJ energies ($\epsilon$)
were used for atoms of different
species~\cite{Jorgensen_DevelopmentTestingOPLS_1996,Jorgensen_Optimizedintermolecularpotential_1984,Watkins_PerfluoroalkanesConformationalAnalysis_2001}.
Non-bonded interactions between atom pairs within a molecule separated
by one or two bonds were excluded. Interaction was reduced by a factor
of $1/2$ for atoms separated by three bonds and more. Simulations were
run using GROMACS version 5~\cite{VanDerSpoel_GROMACSFastflexible_2005}. A
\unit[0.9]{nm} cutoff was employed for the real space part of
electrostatics and Lennard-Jones interactions. The long-range
electrostatics were calculated using particle-mesh Ewald
(PME)~\cite{Essmann_smoothparticlemesh_1995,Darden_ParticlemeshEwald_1993} with the
reciprocal-space interactions evaluated on a 0.16 grid with cubic
interpolation of order 4. First, the system was energy minimized using
the steepest descents algorithm. Then, \unit[10]{ns} simulations in
constant particle number, volume and temperature (NVT) ensemble at
\unit[300]{K} were performed using the stochastic velocity rescaling
thermostat~\cite{Bussi_Canonicalsamplingvelocity_2007} with time constant
\unit[0.1]{ps}. The velocity-Verlet
algorithm~\cite{Verlet_ComputerExperimentsClassical_1967} was employed to integrate the
equations of motions with \unit[2]{fs} time step. The simulation box
size was $\unit[(12\times 12\times 8)]{nm^3}$. Simulations were then
continued in constant particle number, pressure and temperature (NpT)
ensemble at \unit[300]{K} and \unit[1]{bar} controlled by
Parrinello-Rahman~\cite{Parrinello_Polymorphictransitionssingle_1981} barostat with a
coupling time constant of \unit[2.0]{ps}. Molecular visualizations
were done using Visual Molecular Dynamics (VMD)
software~\cite{Humphrey_VMDVisualmolecular_1996}.

\begin{figure}
 \centering
 \includegraphics[width=\linewidth]{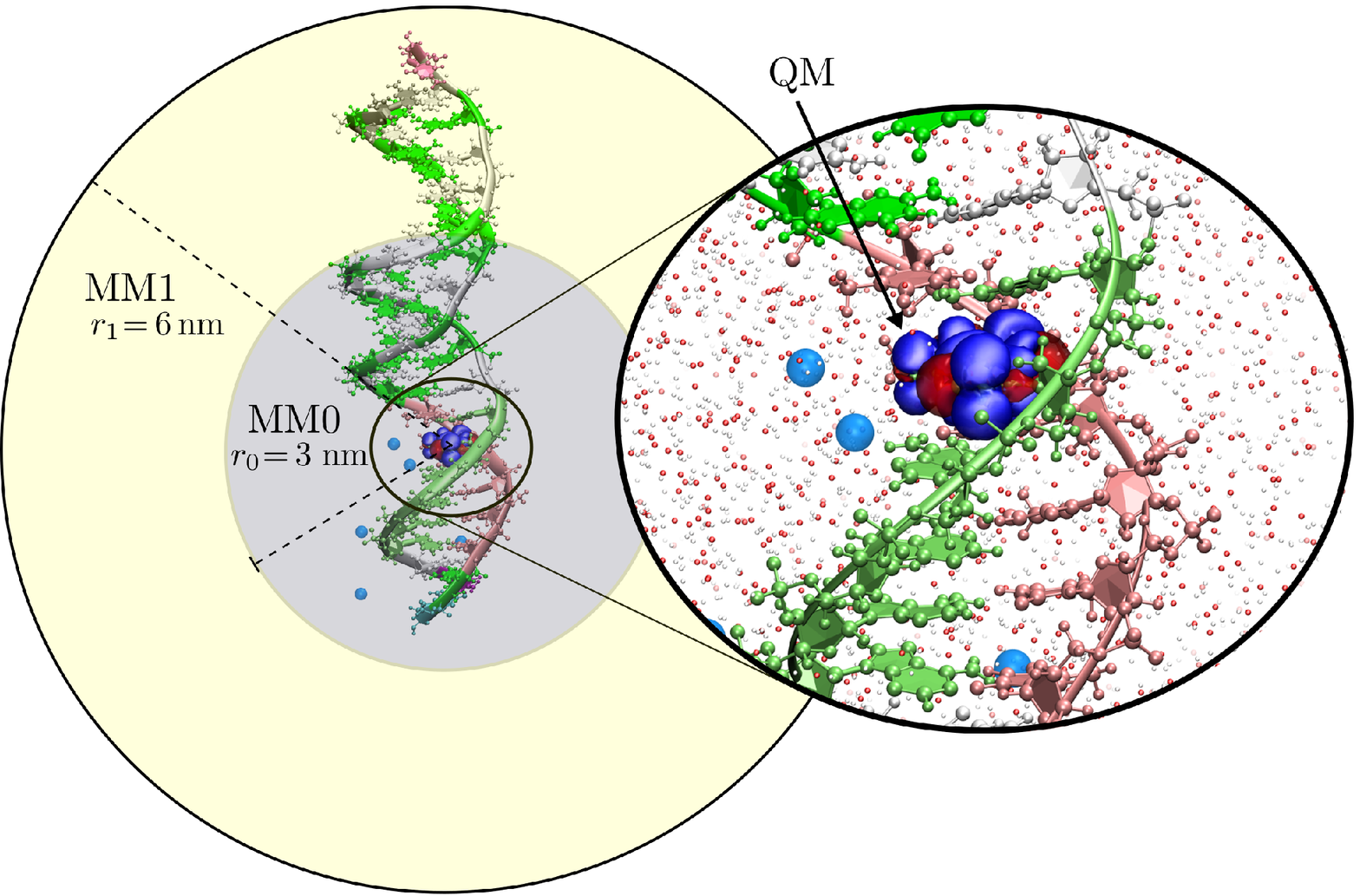}
 \caption{Schematic representation of aqDNA and separation into MM0
   and MM1 for an adenine nucleobase. The QM region is seen in the
   small inset.}
 \label{fig:gwbse:DNA_visual}
\end{figure}

Figure~\ref{fig:gwbse:DNA_visual} illustrates the partitioning of the
MD system of the solvated DNA double strand into QM and MM regions. A
single nucleobase or a base pair is chosen as the QM region, while the
rest of the system that is within a certain distance is assigned to
the MM region. We differentiate between two distinct MM regions, here
referred to MM0 and MM1. In MM0, both static and polarizable effects
are taken into account, while in MM1, only static multipoles are
considered. In this particular case, we restrict the static multipoles
to point charges~\cite{Breneman_Determiningatomcenteredmonopoles_1990} and the induced
moments to dipoles.

For the parametrization of the polarizable model used in the coupled
QM/MM calculations, atomic partial charges and molecular
polarizability tensors were determined for the nucleobases and for
water based on DFT calculations using the PBE0 functional and the
cc-pVTZ basis set. Classical atomic polarizabilities were then
optimized to reproduce the molecular polarizable volume of the DFT
reference calculation. For the DNA backbone, partial charges were
taken from the force field used in the MD simulation and the default
atomic polarizabilities in the AMOEBA
forcefield~\cite{Ponder_CurrentStatusAMOEBA_2010} were employed. Either a single
nucleobase or a pair of nucleobases is chosen as QM region in the
QM/MM setup. As this region is covalently bonded to the MM region, the
bond to the frontier atom was truncated and saturated with a hydrogen
atom. All residues within a closest contact distance of \unit[4.3]{nm}
to the molecules defining the QM region were assigned to the MM
region. When polarized QM/MM calculations were performed, polarization
effects were included for all residues within a closest contact
distance of \unit[2.0]{nm}.

\begin{figure*}
 \centering
 \includegraphics[width=\linewidth]{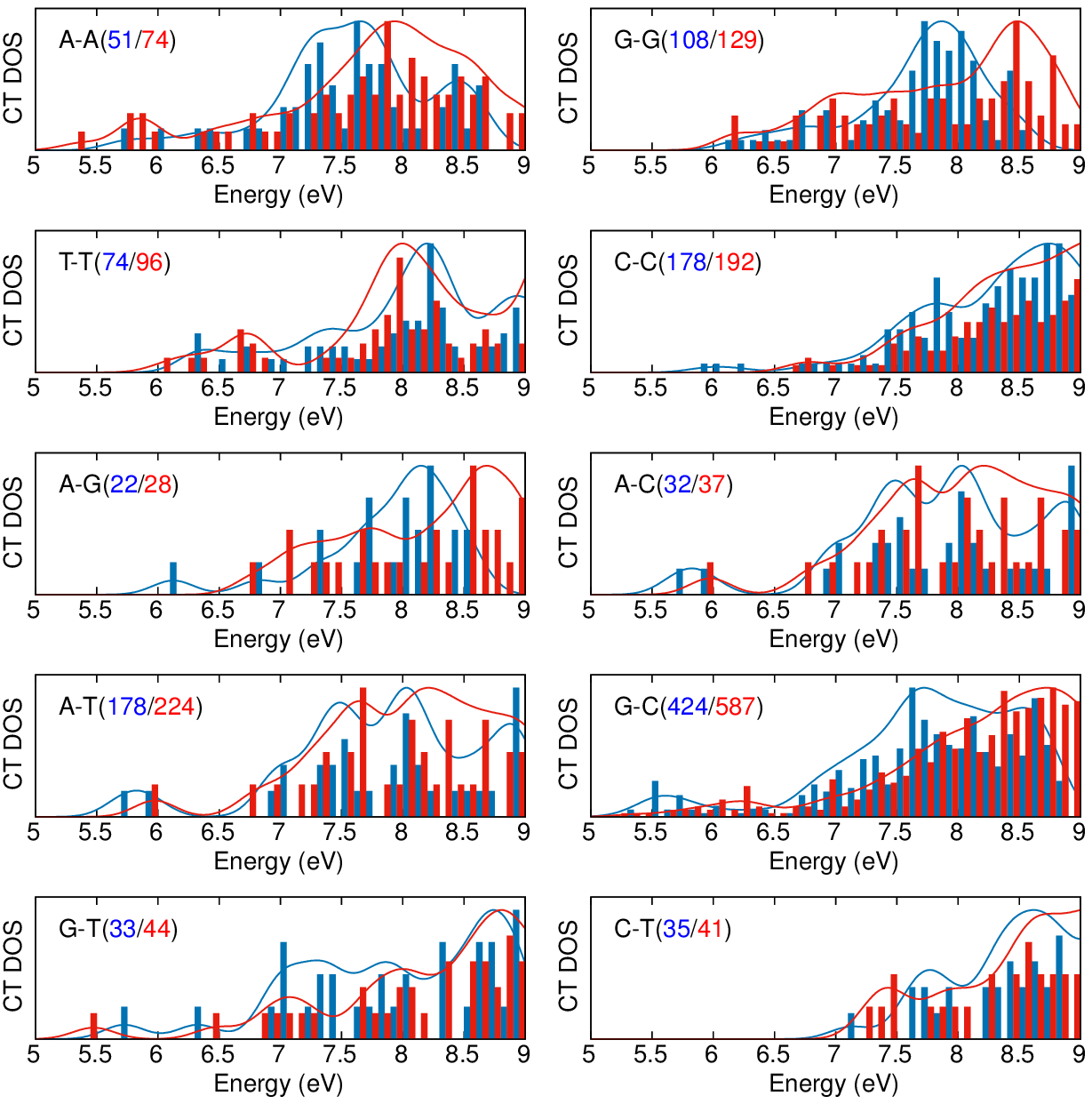}
 \caption{Density of states (DOS) for charge transfer (CT) excitations
   in aqDNA as obtained from dimers in vacuum (blue bars) and QM/MM
   embedded in a static background of point charges (red bars),
   respectively. The individual panels show different base pair
   combinations, in which neighboring nucleobases within a closest
   contact distance of less than \unit[1]{nm} are considered as
   pairs. Due to the specific sequence of the model strand used in
   this work, different numbers of pairs are found for each
   combination. The inset labels indicate both the type of combination
   and in brackets the total number of CT states found in vacuum and
   static QM/MM. A cutoff of \unit[4.3]{nm} was used for the atomistic
   electrostatic embedding.}
 \label{fig:DNA_CTs_basepair}
\end{figure*}

From the simulated DNA structure, neighboring nucleobases with
separation less than \unit[1]{nm} were defined as pairs (yielding 59
pairs in total) between which CT excitations are calculated. These
include both intra- and interstrand excitations. Due to the presence
of the four nucleobases adenine (A), guanine (G), cytosine (C) and
thymine (T) in the present system of aqDNA, 10 different types of
dimers can be formed.

First, we compare the results obtained using QM calculation of
gas-phase dimers with those obtained using QM/MM with only static
classical interactions. Figure~\ref{fig:DNA_CTs_basepair} shows the
distribution of CT exciton energies for both cases. We refer to these
distributions and their Gaussian broadened guide-to-the-eye as CT
Density of States (DOS). CT DOS for dimers in vacuum are respresented
by blue bars, while red bars indicate results for QM/MM dimers
embedded in a static background. The inset labels show the base pair
combination and in brackets the total number of CT states found in
vacuum and static QM/MM, respectively. In all cases, an excitation was
labeled as a CT state if the charge transfer between the two
nucleobases exceeded \unit[0.5]{e}.

A general observation is that the total number of CT states found in
the covered energy region of \unit[5]{eV} to \unit[9]{eV} is always
larger in the QM/MM case. This observation can be attributed to two
effects. First, some of CT states that fall outside of the energy
interval in the gas-phase calculation get pushed down in energy to
values below \unit[9]{eV} in static QM/MM. Second, some of the CT
states change their character by embedding in the static background.

A more detailed analysis of the changes in distributions, in
particular in the low energy regions, show no universal behavior. In
some cases such as for the adenine dimers (A-A) some individual
excitations demonstrate lower energies in static QM/MM than in the
gas-phase. While not resolved in Fig.~\ref{fig:DNA_CTs_basepair}, the
lowest energy CT excitation at about \unit[5.35]{eV} is an intra-stand
adenine dimer of the kind previously discussed by Yin et
al.~\cite{Yin_ChargeTransferExcitedStates_2014} in a more idealized structure. We
will scrutinize the properties of this particular excitation in more
detail below.

In contrast to the behavior of A-A pairs, dimers formed from two
cytosine bases exhibit CT excitons at higher energy than in the
respective gas-phase calculation, irrespective of whether it is an
intra- or interstrand excitation, cf. Figure~\ref{fig:DNA_seqeunce}.

\begin{figure}
 \centering
 \includegraphics[width=\linewidth]{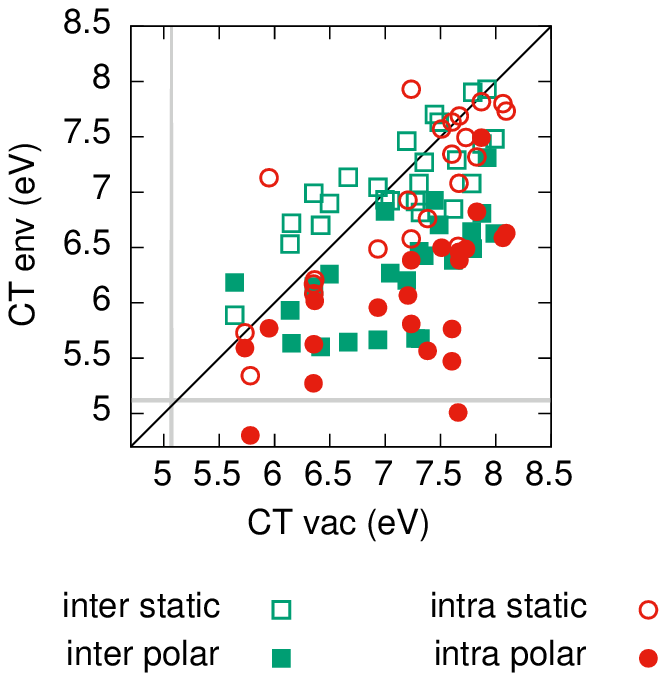}
 \caption{Comparison of CT excitation energies (in eV) calculated in
   static (open symbols) and polarizable (filled symbols) QM/MM setups
   with vacuum QM results. Interstrand (intrastrand) excitations are
   represented by green squares (red circles). The grey shaded areas
   indicate the range of single nucleobase UV absorption energies of
   adenine.}
 \label{fig:DNA_CTs_energies}
 \end{figure}

 Given the non-universal behavior observed upon inclusion of a static
 environment, we limit the following discussion to only the lowest
 energy CT excitation in each of the 59 pairs. The aim is to
 understand the additional influence of polarization in the
 $GW$-BSE/MM calculations. In Figure~\ref{fig:DNA_CTs_energies}, CT
 excitation energies resulting from both static (open symbols) and
 polarized (closed symbols) calculations are shown against the
 respective vacuum energy, also resolving intrastrand (circles) and
 interstrand (squares) excitations. As in the static case, no general
 trend can be discerned. CT excitation energies are both lowered and
 raised due to the presence of the environment. There appears to be
 a tendency that the lower-energy interstrand CTs up to an energy of
 \unit[7]{eV} are all resulting at about \unit[0.5]{eV} higher
 energies in the static case.

 Taking polarization effects into account universally lowers the
 energy, not only with respect to the static QM/MM results, but most
 importantly also with respect to the vacuum calculation. On average,
 we observe a redshift of the interstrand CT energies by
 $\unit[(-0.83\pm0.5)]{eV}$, while intrastrand CTs are redshifted by
 $\unit[(-1.15\pm0.6)]{eV}$, compared to respective vacuum
 results. Notably, these redshifts are on the order of the redshift
 observed in experiment. Also, the CT excitation with the lowest
 energy of \unit[4.81]{eV} is found for a A$_2$ dimer in the chain.

 In addition to the individual CT energies, the grey shaded areas in
 Figure~\ref{fig:DNA_CTs_energies} indicate the energy range in which
 single adenine nuclobases absorb UV light, according to gas-phase and
 QM/MM calculations. While not shown here explicitly, the inclusion of
 a polarizable environment does not affect the energetic properties of
 these localized Frenkel excitons perceptively, with absorption
 predicted to be in the range $\unit[(5.12\pm0.02)]{eV}$. The lowest
 energies of CT excitations found in our dataset are approximately
 \unit[0.3]{eV} below this absorption energy, indicating that the
 decay of the UV excitation to a CT excited state is energetically
 possible, as speculated.

 \begin{figure*}
 \centering
 \includegraphics[width=\linewidth]{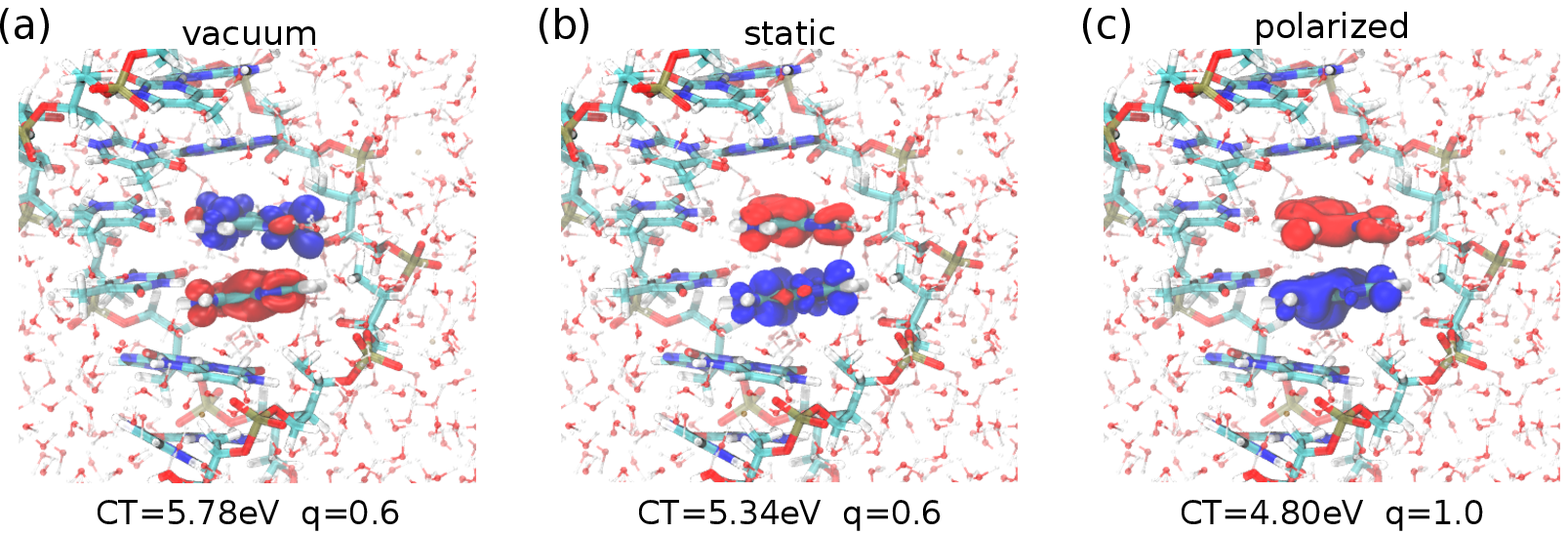}
 \caption{Isosurfaces ($\unit[\pm2 \times 10^{-3}]{e/\text{\AA}^3}$)
   of differential electron densities of the lowest energy adenine dimer
   resulting from (a) a gas-phase (vacuum) calculation, (b) a QM/MM
   calculation with static environment, and (c) a QM/MM calculation
   with polarizable environment. Red color corresponds to negative
   values (hole density) and blue color corresponds to positive values
   (electron density). }
 \label{fig:DNA_CTs_intra}
 \end{figure*}

 Due to this energetic situation, it is worthwhile to analyze the
 A$_2$ CT exciton in further detail and to illustrate how the
 atomistic environment not only affect its energy but also its
 electron-hole wave function. To this end, we show in Figure
 ~\ref{fig:DNA_CTs_intra} the distributions of electron and hole
 densities on the A$_2$ dimer for (a) vacuum QM, (b) static QM/MM, and
 (c) polarized QM/MM, respectively. The associated excitation energies
 and effective charge transfer are indicated below. As discussed
 before, for the vacuum case the CT energy of \unit[5.78]{eV} is
 several \unit[0.1]{eV} above the energy of the UV active
 excitation. The amount of charge transferred in the CT state is only
 \unit[0.6]{e}, with the hole contribution on the lower nucleobase
 (A$_\text{L}$) and the electron contribution on the upper one
 (A$_\text{U}$). Upon inclusion of the static environment, the energy
 of this excitation is lowered by \unit[0.44]{eV} to \unit[5.34]{eV},
 while the amount of charge transferred between the two adenines
 remains at \unit[0.6]{e}. Despite this similarity, the characteristic
 of the excitation is changed significantly, as can be seen in
 Figure~\ref{fig:DNA_CTs_intra}(b). The localization of electron and
 hole contribution in the excitation is inverted. Including
 polarization effects, the general character of the CT excitation
 remains unaffected, i.e., the hole is localized on A$_\text{U}$ and
 the electron on A$_\text{L}$. Most notably, however, the excitation
 exhibits integer charge transfer character in this situation.

 \begin{figure}
 \centering
 \includegraphics[width=\linewidth]{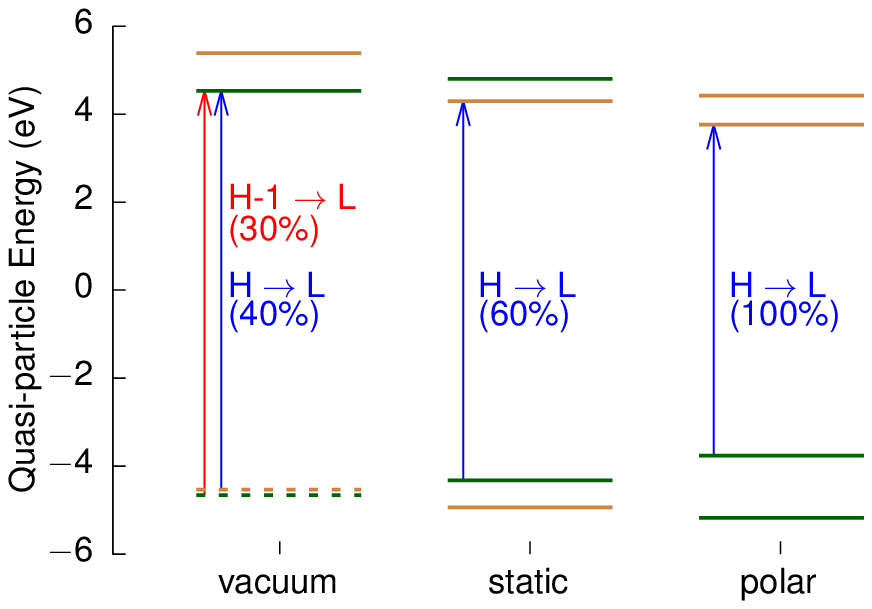}
 \caption{Quasi-particle energy levels (eV) for HOMO-1, HOMO, LUMO,
   and LUMO+1 resulting from (a) a gas-phase (vacuum) calculation, (b)
   a QM/MM calculation with static environment, and (c) a QM/MM
   calculation with polarizable environment. The color of horizontal
   lines indicates the localization of the quasi-particle states on
   the either of the two nucleobases. Brown (dark green) represents
   localization on A$_\text{L}$ (A$_\text{U}$). For HOMO-1 and HOMO in
   the vacuum case, the quasi-particle states are distributed over the
   whole base pair, which is noted as a dashed line. Vertical arrows
   show the dominant transitions forming the CT excitation.  }
 \label{fig:DNA_CTs_QP}
 \end{figure}

\begin{figure}
 \centering
 \includegraphics[width=\linewidth]{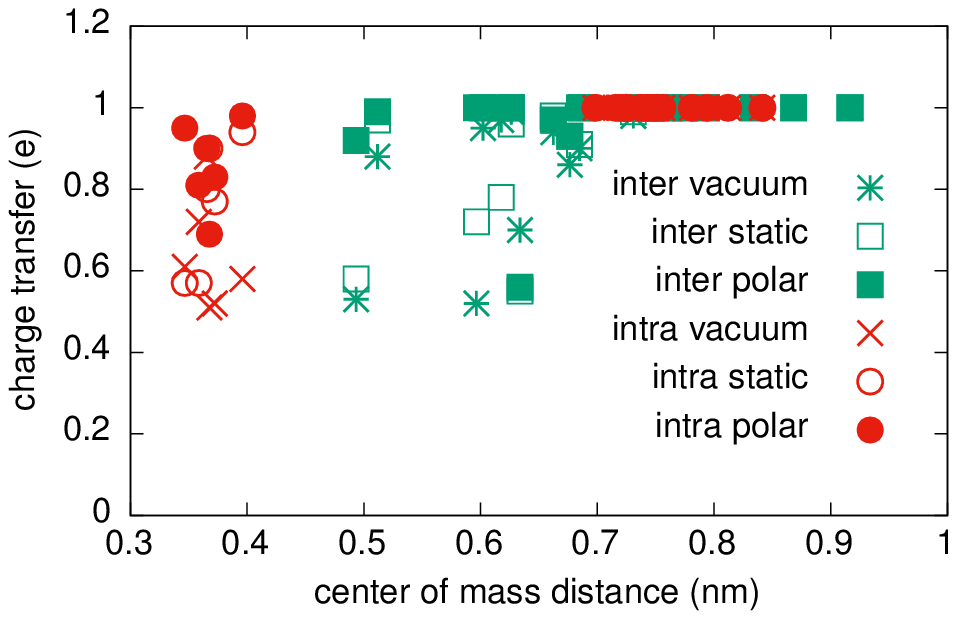}
 \caption{Effective charge transfer character (in $e$) in the CT
   excitations as a function of center-of-mass distance of the
   involved monomers (in nm). Results for intra- and interstrand
   excitations are compared for the three different calculation
   setups: vacuum, static QM/MM, and polarized QM/MM. }
 \label{fig:DNA_CTs_CT}
\end{figure}

 The observation that the nature of the CT excitation can be affected
 dramatically by the complex molecular environment can be attributed
 to a combination of a shift of energy levels and changed composition
 of transitions. We analyze the quasi-particle energy levels obtained
 at the $GW$ step of the respective
 calculations. Figure~\ref{fig:DNA_CTs_QP} shows the energies of two
 highest occupied and two lowest empty quasi-particle levels for
 vacuum, static, and polarized calculations. Note that for an easier
 comparison, the zero of the energy scale has been set to the center
 of the HOMO-LUMO gap in all individual cases. The spatial
 distribution of all quasi-particle wave functions has been inspected
 and is indicated by the horizontal lines' color. Brown (dark green)
 lines indicate states that are localized on A$_\text{L}$
 (A$_\text{U}$). In addition, the vertical lines show the
 contributions of the quasi-particle transitions to the respective CT
 excitations in Figure~\ref{fig:DNA_CTs_CT}, with the weights given in
 the inset.

 In the case of the vacuum calculation on the adenine dimer taken from
 the MD snapshot, it turns out that the two occupied levels cannot be
 uniquely assigned to either of the two nucleobases. Instead, the
 quasi-particle states delocalize over the dimer, however not at equal
 distribution. Note, though, that they are only separated by
 \unit[0.13]{eV} in energy. To make this also visually clear, the two
 levels are shown as dashed lines in Figure~\ref{fig:DNA_CTs_QP}. As
 can be seen from the two arrows, the CT excitation in this
 environment-free QM calculation is composed of HOMO-1 $\to$ LUMO and
 HOMO $\to$LUMO transitions with nearly equal weight. The fact that
 the combined weight is only \unit[70]{\%} emphasizes that even more
 quasi-particle transitions play a significant role here. It is likely
 that this is directly linked to the delocalized nature of the
 occupied states. Taken as a whole, the hole contribution of the CT,
 arising in large parts from the HOMO and HOMO-1 states, is
 consequently localized on A$_\text{L}$. For the two unoccupied levels
 shown here, no strong delocalization over the dimer can be
 identified. Since the LUMO is localized on A$_\text{U}$, also the
 electron density in the CT state is found on this nucleobase.

 Turning now towards the results obtained from calculation performed
 in the static QM/MM setup, one can spot significant changes as
 compared to the vacuum only calculation. First, all quasi-particle
 states around the HOMO-LUMO gap are localized on either of the two
 nucleobases of the excimer. In the occupied manifold, one can now
 assign the HOMO to be uniquely localized on A$_\text{U}$ and HOMO-1
 on A$_\text{L}$. As a consequence the energetic separation is more
 pronounced, amounting to \unit[0.62]{eV}. At the same time also the
 two unoccupied states change character. While also localized on
 either of the two nucleobases in the vacuum calculation, one finds
 that the specific localization site is switched. The LUMO is now
 localized on A$_\text{L}$, and LUMO+1 on A$_\text{U}$. Combined with
 the fact that the dominant transition in the CT excitation is a HOMO
 to LUMO transition from A$_\text{U}$ to A$_\text{L}$ with a weight of
 approximately \unit[60]{\%}, see Figure~\ref{fig:DNA_CTs_QP}, the
 localization behavior of hole and electron densities is inverted as
 compared to the vacuum case. The total transferred charge however
 remains \unit[0.6]{eV}, which can be attributed to the
 additional transitions that collectively contribute to \unit[40]{\%}
 of the excited state.

 We note in passing that the HOMO-LUMO gap is also slightly reduced by
 the embedding in a static molecular environment, namely from
 \unit[9.07]{eV} to \unit[8.64]{eV}. A word of caution: The fact that
 the reduction by \unit[0.43]{eV} of this gap is numerically similar
 to the lowering of the CT excitation energy by \unit[0.44]{eV} is
 likely coincidental. Typically, a change in localization of the
 contribution quasi-particle states leads to a very different
 composition of the effective electron-hole interaction that
 determines the exciton binding energy and, concomitantly, the
 excitation energy.

 From the quasi-particle levels in the polarized QM/MM calculation as
 shown on the right-hand side of Figure~\ref{fig:DNA_CTs_QP}, one can
 see that the environment polarization response modifies this picture
 even more. First of all, now one finds the two occupied states shown
 being localized on the upper adenine nucleobase, and the two
 unoccupied states localized on the lower one. The HOMO-LUMO gap is
 further reduced to \unit[7.52]{eV}, and the energetic separation of
 the occupied and unoccupied levels is increased. Most remarkable is
 now that the CT excitation is in this case given as a pure HOMO to
 LUMO transition. The hole and electron contributions are fully
 localized on A$_\text{U}$ and A$_\text{L}$, respectively,
 corresponding to integer charge transfer.

 The above detailed analysis of the characteristics of the
 quasi-particle and CT excited states for the minimum energy CT found
 in our data set clearly reveals that the resulting excitation
 energies in complex molecular environments obtained from QM/MM
 calculations are a result of an intricate interplay of several
 effects. In particular, modifications on the nature of the
 quasi-particle states are significant since their
 localization/delocaliztion characteristic have a profound and direct
 effect on the two-particle excitations. This interplay cannot be
 captured by adding a perturbative energy correction due
 to the environment to a vacuum QM calculation.

To scrutinize whether the change of effective charge transfer in the
CT excitation observed for the intra-strand adenine dimer observed
above is a more general effect of embedding into a static and/or
polarizable molecular environment, we show in
Figure~\ref{fig:DNA_CTs_CT} the calculated amount of transferred
charge as a function of center-of-mass distance for the various
calculation setups. We differentiate also between intra- and
inter-strand excitations.

It can be seen for the excimers with the closest intermolecular
separation between \unit[3]{\AA} and \unit[4]{\AA}, which are
exclusively intra-strand excitations, vacuum calculations yield only
partial charge transfer upon excitation between \unit[0.5]{$e$} and
\unit[0.9]{$e$}. The same holds for inter-strand dimers with distances
of \unit[5]{\AA} and \unit[6]{\AA}. All these short distance dimers
are essentially neighboring molecules whose electron density can
spatially overlap and the associated interaction yielding (partially)
delocalized quasi-particle states. For all dimers with separation
larger than \unit[0.7]{nm} center-of-mass distance, i.e.,
second-nearest neighbors, such a direct interaction is not
possible. In case of intra-strand excitations, it means that in a
stack of three bases (base trimer), only the outer two nucleobases are
treated quantum-mechanically, while the center one is part of the
polarizable MM region. This is strictly speaking a fairly strong
approximation. When base stacking interactions are strong, the purely
classical treatment cannot cover possible effects of forming
delocalized states and the associated partial charge transfer. Also,
such explicit base pair interactions might affect the CT excitation
energies directly. A possible pathway to cover such effects is to
treat the full base trimer quantum-mechanically and embed this in a
classical environment. However, this case goes beyond the scope of
this work and is left for future studies.

We focus in the following on the short-distance excimers. When the
molecular environment is taken account, the static-only interactions
(open symbols in Figure~\ref{fig:DNA_CTs_CT}) affect the amount of
effectively transferred charge roughly in the same fashion as observed
for the A$_2$ system with minimal CT excitation energy discussed
above. In some cases, one can note a change of this effective charge
by up to \unit[0.3]{e}. However, at least for the first shell of
intra- and inter-strand dimers, there is no observable integer charge
transfer state.

Only upon adding environment polarization effects (filled symbols in
Figure~\ref{fig:DNA_CTs_CT}, most of the CT states are approaching
such an integer CT character. It stands to reason that remnant
delocalization for quasi-particle states is responsible for that.


\section{Summary}

\label{sec:summary}
In this paper, the Gaussian-orbital based implementation of many-body
Green's functions theory within the $GW$ approximation and the
Bethe-Salpeter Equation (BSE) in the open-source \xtp software has
been introduced. Application to the standard small molecule Thiel set
has been used to benchmark the obtained excitation energies. The
results are in very good agreement with the experimental reference for
a variety of excitation types and an energy range from
\unit[2-8]{eV}, validating both the methodology and its implementation.

It has further been demonstrated how coupling $GW$-BSE to a classical
atomistic environment in QM/MM schemes allows studying electronic
excitations in complex molecular environments, here in prototypical
aqueous DNA. It is found that charge transfer excitations are
extremely sensitive to the specific environment. For the lowest energy
CT excitations in an intrastrand adenine dimer, the approach predicts
energies below that of the UV active single nucleobase
excitation. This has a tremendous impact on the possibility of an
inital (fast) decay of such an UV excited state into a bi-nucleobase
CT exciton, which is considered one the pathways for UV-induced
DNA damage. The calculated redshift of the CT excitation energy
compared to a nucelobase dimer treated only in vacuum is of the order
of \unit[1]{eV}, which matches expectations from experimental data.
The $GW$-BSE/MM methodology used here allows to gain very
detailed insight into the mechanisms leading to the observed
energies. It is possible to disentangle the effects of the different
levels of the explicit molecular environment on single-particle and
two-particle excitations. Incorporating $GW$-BSE into the presented
QM/MM setup is therefore an extremely powerful tool to study a wide
range of types of electronic excitations in complex molecular
environments.

\section*{Acknowledgements}
This work has been supported by the Innovational Research Incentives
Scheme Vidi of the Netherlands Organisation for Scientific Research
(NWO) with project number 723.016.002. We gratefully acknowledge the
support of NVIDIA Corporation with the donation of the Titan X Pascal
GPU used for this research.

\section*{Supporting Information Available}
The Supporting Information consists of a PDF document containing
tables with the size of the basis sets used, a comparison between
results obtained with $G_0W_0$ and $\text{eV}GW$ and DFT functional
dependence, as well as the full excitation data for the Thiel
set. Figures show (i) the convergence of excitation energies with the
number of levels taken into account in the RPA, (ii) a comparison of our
$GW$-BSE results to TBE instead of experiment, and (iii) a comparison
of full BSE results with the use of the Tamm-Dancoff Approximation.

\bibliography{literature_short}

\end{document}